\documentstyle[12pt]{article}
\title{On reduction of the wave-packet, decoherence, irreversibility and the second law of thermodynamics}
\author{H. Narnhofer\\
        Institut f\"{u}r Theoretische Physik\\
        Universit\"{a}t Wien\\
        Boltzmanngasse 5\\
        A1090 Vienna, Austria\\
        \texttt{heide.narnhofer@univie.ac.at}
        \and
        W. F. Wreszinski\\
        Instituto de Fisica\\
        Universidade de S\~{a}o Paulo\\
        Rua do Mat\~{a}o, Travessa R 187\\
        05508-090 S\~{a}o Paulo, Brazil\\
        \texttt{wreszins@gmail.com}
        }
\begin{document}
\maketitle
\begin{abstract}
We prove a quantum version of the second law of thermodynamics: the (quantum) Boltzmann entropy increases if the initial (zero time) density matrix decoheres, a condition generally satisfied in Nature. It is illustrated by a model of wave-packet reduction, the Coleman-Hepp model, along the framework introduced by Sewell (2005) in his approach to the quantum measurement problem. Further models illustrate the monotonic-versus-non-monotonic behavior of the quantum Boltzmann entropy in time. As a last closely related topic, decoherence, which was shown by Narnhofer and Thirring (1999) to enforce macroscopic purity in the case of quantum K systems, is analysed within a different class of quantum chaotic systems, viz. the quantum Anosov models as defined by Emch, Narnhofer, Sewell and Thirring (1994).

A review of the concept of quantum Boltzmann entropy, as well as of some of the rigorous approaches to the quantum measurement problem within the framework of Schr\"{o}dinger dynamics, is given, together with an overview of the C* algebra approach, which encompasses the relevant notions and definitions in a comprehensive way.
\end{abstract}


\tableofcontents

\section{Introduction and summary}

\subsection{Quantum theory of measurement and irreversibility}

The subject of quantum measurement theory and the related problems of irreversibility and decoherence has a long history, beautifully reviewed and summarized in \cite{Wightman1},\cite{Wightman2}. We refer to the classic work of von Neumann \cite{vN}, as well as to the ''big red book'', as Wightman calls it \cite{WhZu}, as early references, and to the review articles of Zurek \cite{Zu1}, \cite{Zu2}, as well as the books \cite{Pe}, \cite{Giu} \cite{Schlossauer}, and the recent article \cite{LandReu} for comprehensive treatments of the subject and references.

The recent review by Allahverdyan, Balian and Nieuwenhuizen \cite{ABN1} complements our treatment nicely, both from the conceptual point of view (the central concepts we use, the quantum Boltzmann entropy and quantum Anosov systems, are not contemplated in \cite{ABN1}) and the models which are analysed. Indeed, the authors of \cite{ABN1} treat most extensively a mean field model, of interaction between Curie-Weiss spins and a phonon bath (see also \cite{ABN2}), while in our applications in chapter 5 we study models of local interaction between a particle and a spin chain. 

On the other hand, rigorous treatments of the quantum theory of measurement within the framework of Schr\"{o}dinger dynamics, which we adopt, have been extremely rare. We are only aware of the papers of Hepp \cite{He}, Sewell (\cite{Sew1}, \cite{Sew2}), Whitten-Wolfe and Emch \cite{WWEm}, see also \cite{Land} for a comprehensive review. A recent paper on irreversible behavior and collapse of the wave-packets in a quantum system with point interactions is \cite{Gua}: the model lacks, however, for the moment, a physical interpretation.

In section 3 we provide a review of the approaches of \cite{He} and (\cite{Sew1}, \cite{Sew2}) to the quantum measurement problem. Since we hope to make the present paper accessible to theoreticians, not only mathematical physicists or mathematicians, in section 2 an overview of what is needed from the C* algebra approach of infinite quantum systems is given (for a more comprehensive account, see \cite{Land} and \cite{Se3}), as well as \cite{Hug}).

In section 4 we discuss the problem of irreversibility and its connection to the quantum measurement problem. The quantum Boltzmann approach has been discussed by C. Maes and coworkers in a similar spirit, albeit different direction: see \cite{Maes2} for a comprehensive exposition and references. In particular, the model \cite{deRJMN} is an example of a unitary quantum dynamics with positive entropy production. The subject has also been recently illuminated in a thorough and precise discussion of quantum probability by J Fr\"{o}hlich and B. Schubnel \cite{FrSch}. In particular, the operator algebra viewpoint is treated in greater depth there, and a unified discussion of concepts such as decoherence, information loss and entanglement is provided.

A central object in our approach is what we call the quantum Boltmann entropy. In order to explain its main features, it is convenient to consider classical systems first, for which a comprehensive theory of the approach to equilibrium, characterized by the growth of the Boltzmann entropy to a limiting value, exists (\cite{GLeb}, see also \cite{Leb}). 

\subsection{The classical Boltzmann entropy}

We now follow \cite{Leb} quite closely, for the reader's convenience, because Lebowitz's discussion contains all the essential points.

The microscopic state of a classical system of $N$ particles is represented by a point
$X \equiv (\vec{r_{1}},\vec{p_{1}}, \cdots , \vec{r_{N}},\vec{p_{N}})$ in its phase space $\Gamma$, $\vec{r_{i}}$ and
$\vec{p_{i}}$, $i=1, \cdots, N$ denoting the position and momentum of the ith particle. We suppose the dynamics to be Hamiltonian, associated to a flux $T_{t}$ which is phase-space volume preserving, by Liouville's theorem, describing the macroscopic state. The latter will be denoted by $M$ and consists of a system of $N$ (of order of Avogadro's number) atoms in a box $V$, with volume $|V| \ge Nl^{3}$, where $l$ is a typical atomic scale. Divide $V$ into $K$ ''cells'', where $K$ is large but still $K \ll N$, and specify number of particles, momentum and energy in each cell. All assertions in the following are to be taken with some tolerance, one thinks of the ''thickened'' microcanonical ensemble, see, e.g., \cite{Pen}, pg. 1960. $M$ is thus determined by $X$ - we write, always following \cite{Leb}, $M(X)$, but there is a continuum of $X$ corresponding to the same $M$. Let $\Gamma_{M}$ denote the region in $\Gamma$ consisting of all microstates $X$ corresponding to a given macrostate $M$, and $|\Gamma_{M}| = \frac{\int_{\Gamma_{M}}\prod_{i=1}^{N} d\vec{r_{i}}d\vec{p_{i}}}{N!h^{3N}}$.

Let, now, a gas of $N$ atoms with energy $E$ be initially confined by a partition to the left half of the box $V$, and suppose that this constraint is removed at time $t_{0}$. The phase space volume available to the system for times $t > t_{0}$ is, then, an enormous factor $2^{N}$ of the initial one. The space of macrostates consists in this case of all couples of pairs $\{(N_{L},E_{L}),(N_{R},E_{R})\}$ satisfying both constraints $N_{L}+N_{R}=N$ and $E_{L}+E_{R}=E$.  Equivalently, we may label a macrostate of this gas $M = (\frac{N_{L}}{N},\frac{E_{L}}{E})$, the fraction of the particles and energy in the left half of $V$; the macrostate $M_{0}$ at time $t_{0}$ is $M_{0}=(1,1)$. The totality of all phase points $X_{t}$, which at time $t_{0}$ were in $M_{0}$, forms a region $T_{t}\Gamma_{M_{0}}$, whose volume is equal to $|\Gamma_{M_{0}}|$, but whose \textbf{shape} will increasingly be contained in regions $\Gamma_{M}$ corresponding to macrostates with larger and larger phase space volumes, until almost all the phase points initially in $\Gamma_{M_{0}}$ are contained in $\Gamma_{M_{eq}}$, where $M_{eq}=(1/2,1/2)$ (up to fluctuations, see \cite{Leb}). Boltzmann assumed that $S_{B}(M_{eq}) \equiv k\log|\Gamma_{M_{eq}}|$ is proportional to the thermodynamic entropy of Clausius, and generalised it to define an entropy also for macroscopic systems not in equilibrium, by associating with each microscopic state $X$ the number
$$
S_{B}(M(X)) \equiv k \log|\Gamma_{M(X)}|
\eqno{(1.1)}
$$
which, following Penrose \cite{Pen1}, Lebowitz calls the Boltzmann entropy of a classical system. Boltzmann then argued that, due to the large differences in the sizes of $\Gamma_{M}$, $S_{B}(X_{t}) = k\log|\Gamma_{M(X_{t})}|$ \textbf{typically} increases in a way which qualitatively describes the evolution towards equilibrium of macroscopic systems. ''Typically'' means that there will be ''bad'' intitial microstates $X \in \Gamma_{M_{0}}$ - e.g., those which have all velocities directed away from the barrier lifted at $t_{0}$ in the gas example - but the phase space volume of these is an utterly negligible fraction of $|\Gamma_{M_{0}}|$.

We are here speaking of \textbf{closed} systems - systems completely isolated from all external influences. Traditional thermodynamics takes this idealization as the starting point - see Clausius' sweeping formulations of the first and second laws, ''Die Energie der Welt ist konstant'' and ''Die Entropie der Welt strebt einem Maximum zu'' (\cite{tHW}: ''The energy of the world is constant'' and ''The entropy of the world tends to a maximum''). Accordingly, we take closed systems as our fundamental objects, hoping, as in \cite{Gr} that, once irreversibility is understood for closed systems, ''it is possible to make some estimate of the severity of the approximations involved''. Although there has been great progress in the theory of irreversibility for open quantum systems (see \cite{JPi1},\cite{JPi2} and references given there) examples such as the evolution of the Universe or the adiabatic irreversible expansion of a gas, in which the isolation of the system may be achieved to any degree of accuracy - a prototype of events of daily experience - demonstrate the necessity of considering both open and closed systems. The procedure in thermodynamics \cite{LYPR} is also in conformance to our approach; compare also the discussion in (\cite{Haag}, first paragraph page 311).

\subsection{Difficulties with the quantum version: macrostates, subsystems, Schr\"{o}dinger cats and entanglement}

For quantum systems, a macrostate $M$ of a macroscopic quantum system is described by von Neumann \cite{vN} (see the discussion in \cite{Leb}, section 8, which we follow below) by specifying the values of a set of commuting macroscopic observables $\hat{M}$, representing particle number, energy, etc., in each of the cells into which the box containing the system is divided. Labelling the set of eigenvalues of $\hat{M}$ by $M_{\alpha}$, with $\alpha =1, \cdots, n$, we have, corresponding to the collection $\{M_{\alpha}\}$ an orthogonal decomposition of the system's Hilbert space ${\cal H}$ into linear subspaces $\hat{\Gamma_{\alpha}}$ in which the observables assume the values $M_{\alpha}$. Let $P_{\alpha}$ denote the projection onto $\hat{\Gamma_{\alpha}}$, $\hat{\mu}$ the density matrix describing the system ($\hat{\mu} = |\Psi)(\Psi|$ for a system in a pure state associated to a wave function $\Psi$), 
$$
p_{\alpha} = Tr(P_{\alpha}\hat{\mu})
\eqno{(1.2)}
$$
and
$$
\tilde{\mu} = \sum (\frac{p_{\alpha}}{|\hat{\Gamma_{\alpha}}|}) P_{\alpha}
\eqno{(1.3)}
$$
The macroscopic von Neumann entropy $S_{mac}(\hat{\mu})$ is then defined by
$$
S_{mac}(\hat{\mu}) \equiv -kTr(\tilde{\mu}\log\tilde{\mu})
\eqno{(1.4)}
$$
Let $\{M_{\alpha}\}$ denote the set of possible macrostates of a classical system. It happens, however, that, whereas in the classical case the actual microstate of the system is described by $X \in \Gamma_{M_{\alpha}}$, for some $\alpha$, so that the system is definitely in one of the macrostates $M_{\alpha}$, this is not so for a quantum system specified by $\hat{\mu}$ or $\Psi$, i.e., the analogue of (1.1) is not valid. Indeed, even when the system is in a microstate $\Psi$ or $\hat{\mu}$, corresponding to a definite macrostate at time $t_{0}$, only a classical system will be in a unique macrostate for all times $t$, because the quantum system may evolve to a superposition of different macrostates (''Schr\"{o}dinger cats''), an example of which appears in section 5.1. In addition, and related to this fact, is that, in contrast to the classical case, the state of a subsystem (''cell'') depends on possible correlations to other systems - i.e., if one extends the algebra of observables, different (e.g. ''entangled'') states may be obtained. As Landsman aptly remarks \cite{Land}, ''the restrictions of a generic state to all subsystems of a system does not, in general, uniquely determine the state on the total system''. This means, grosso modo, that it is not possible to assign a wave-function to a subsystem - the analogue of the previous ''cells'' in classical statistical mechanics.

Both phenomena discussed above - Schr\"{o}dinger cats and entanglement (in this connection see \cite{Zu3})- arise in quantum measurement theory. It seems therefore of special interest to study irreversibility in the quantum theory of measurement. This is attempted in sections 4 and 5.

In order to do so, one really needs a formulation and proof of a form of the second law of thermodynamics for closed quantum systems, which includes both the measurement and the thermal cases. This is done in section 4. 

\subsection{The second law for closed quantum systems and the assumption of decoherence at zero time for the initial state}

Following Lindblad \cite{Lind1}, we define a quantum Boltzmann entropy $S_{QB}$ (related but not identical to
$S_{mac}$) and show that for automorphic (unitary) time evolutions, with a initial state $\hat{\mu}=\hat{\mu}(0)$ which is decoherent at time zero (definition 4),
$$
S_{QB}(0) \le S_{QB}(t)
\eqno{(1.5)}
$$
for all $t \ge 0$ (theorem 1). The growth in (1.5) is, however, not necessarily monotonic (see remark 1), and for this reason we speak of two fixed states, an initial state out of equilibrium satisfying the above assumption and a final state, which may be a stationary and/or equilibrium state. In section 5 we illustrate theorem 1 by a model of a quantum measurement - the Coleman-Hepp model, along the framework introduced by Sewell and reviewed in 3.5. Further models in section 5 (section 5.8) investigate the important issue of the monotonic- versus- non-monotonic time behavior of the quantum Boltzmann entropy.

The assumption of decoherence at zero time for the initial state (density matrix) is related to the property of macroscopic purity for systems with infinite number of degrees of freedom, which is found in Nature. Macroscopic purification may be obtained by successive interactions with the environment, if the latter is represented by a so-called quantum K system (\cite{NTh2},\cite{Em}). This fact was proved by Narnhofer and Thirring \cite{NTh1}. Quantum K system are at the top of the hierarchy of quantum chaotic systems. This led us to investigate the closely related problem of decoherence for a different class of quantum chaotic systems, the quantum Anosov systems, introduced by Emch,Narnhofer, Sewell and Thirring \cite{ENST} in section 5.9, which allow some explicit estimates on the decoherence time.

In section 5 we consider examples and partly restrict ourselves to the case of finite-dimensional Hilbert space. However, even for finite number of degrees of freedom, it will be necessary to consider infinite-dimensional spaces, and the related delicate problems (see, e.g., the appendix by B. Simon to \cite{LRu}) have not been considered, as well as the thermodynamic limit of the specific entropy (\cite{LRu}, see also\cite{ArSe}) .

Section 6 is reserved to the conclusion.

\section{A short overview of the C* algebra approach}

\subsection{General notions: algebras, states, representations; disjointness and coherence; folia and superselection sectors.}

We give a preliminary overview of the $C^*$ algebra approach to infinite quantum systems as they are used in Hepp's approach.
According to the Haag-Kastler theory of local algebras of observables \cite{HK}, one assumes that for each bounded region
$\Lambda \subset \mathbf{R}^{3}$ ($\Lambda \subset \mathbf{Z}^{3}$ for a lattice system) there is an associated algebra
${\cal A}(\Lambda)$, and $\Lambda_{1} \subset \Lambda_{2}$ implies ${\cal A}(\Lambda_{1}) \subset {\cal A}(\Lambda_{2})$. By taking the union over all $\Lambda$ the \textbf{local algebra} ${\cal A}_{L} \equiv \cup_{\Lambda}{\cal A}(\Lambda)$ is obtained. Haag and Kastler assumed that ${\cal A}(\Lambda)$ are C* algebras, which have a norm $||.||$ and are complete in the topology defined by the norm, i.e., if $\{A_{n}\}_{n=1}^{\infty} \in {\cal A}_{L}$ and $\lim_{n \to \infty}||A-A_{n}||=0$, then $A \in {\cal A} \equiv \overline{{\cal A}_{L}}$, where the bar denotes the above described norm closure: ${\cal A}$ is called the \textbf{quasi-local algebra}. This is a way to arrive at an infinite system: although the limit points of
${\cal A}$ do not belong to the ${\cal A}(\Lambda)$ of any bounded region $\Lambda$, they may be viewed as ''essentially local''.

A C* algebra ${\cal A}$ is, thus, a vector space over the complex numbers, with an associative multiplication operation compatible with the vector space structure, an antilinear involution $\ast$, satifying $A^{\star\star}=A$ and
$(AB)^{\star}=B^{\star}A^{\star}$ for all $A,B \in {\cal A}$, a norm satisfying the triangle inequality, as well as the relation $||AB|| \le ||A||||B||$, in such a way that ${\cal A}$ is complete in the topology induced by the norm, and, finally, the equality $||A^{\star}A||=||A||^{2}$. This definition may be motivated by the fact that it is satisfied by norm-closed subalgebras of ${\cal B}({\cal H})$, the set of bounded operators on a separable Hilbert space ${\cal H}$, with
$||A||^{2} = \sup_{\Phi}(\Phi, A^{\star}A \Phi)$, with the supremum taken over all unit vectors $\Phi \in {\cal H}$, where, above, $\star$ is the usual adjoint operation, and $(.,.)$ denotes the inner product on ${\cal H}$, antilinear in the first argument. One simple example of the latter structure is, on ${\cal H} = \mathbf{C}^{2}$, given by the Pauli matrices
$\sigma^{\alpha},\alpha=1,2,3$ satisfying the usual algebra
$$
\sigma^{1}\sigma^{2}=i\sigma^{3}
\eqno{(2.1)}
$$
plus cyclic permutations, with the adjunction of a unit $\mathbf{1}$. ${\cal A}_{\Lambda}$, for $\Lambda \subset \mathbf{Z}^{d}$, for instance ($d \ge 1$; in the example of section 3.1, d=1), becomes the concrete spin algebra on
${\cal H}(\Lambda)= \otimes_{i\in \Lambda}\mathbf{C}^{2}_{i}$, generated by taking linear combinations, (ordinary and tensor) products of $\sigma^{j}_{i} \mbox{ and } \mathbf{1}_{i} \mbox{ with } i \in \Lambda \mbox{ and } j=1,2,3$. The local algebra is
$$
{\cal A}_{L} =\cup {\cal A}(\Lambda)
\eqno{(2.2.1)}
$$
with the quasilocal algebra
$$
{\cal A} = \overline{{\cal A}_{L}}
\eqno{(2.2.2)}
$$
which will be called the spin algebra and denoted ${\cal A}_{spin}$.

For particles, e.g. in one dimension, one considers
${\cal H}=L^{2}(\mathbf{R})$ on which the concrete algebra ${\cal W}$ of finite linear combinations of (bounded) Weyl operators
$$
W(\beta,\gamma) = \exp[i(\beta x+\gamma p)]
\eqno{(2.3.1)}
$$
acts. These operators satisfy the Weyl form of the canonical commutation relations:
$$
W(\beta,\gamma)^{\dagger} = W(-\beta,-\gamma)
\eqno{(2.4.1)}
$$
together with
$$
W(\beta,\gamma)W(\beta^{'},\gamma^{'})=\exp[-1/2(\beta\gamma^{'}-\beta^{'}\gamma)]W(\beta+\beta^{'},\gamma+\gamma^{'})
\eqno{(2.4.2)}
$$
Their norm is given by
$$ ||W(\beta,\gamma)||=1 \quad ||W(\beta,\gamma)-W(\beta ',\gamma ')||=2, \quad \beta \neq \beta ' ;\gamma \neq \gamma '
\eqno{(2.4.3)}
$$
A \textbf{state} on a C* algebra ${\cal A}$ is defined to be a positive linear form $\omega$ on ${\cal A}$, that is normalized to one on the unit element $\mathbf{1}$ of ${\cal A}$:
\begin{eqnarray*}
\omega(\alpha A + \beta B) = \alpha \omega(A) + \beta \omega(B)\\
\omega(A^{\star}A) \ge 0\\
\omega(\mathbf{1}) = 1
\end{eqnarray*}
We shall see examples very shortly. A $\star$-representation of ${\cal A}$ on a Hilbert space ${\cal H}$ is a mapping $\Pi$ of ${\cal A}$ into ${\cal B}({\cal H})$, such that
\begin{eqnarray*}
\Pi(\alpha A + \beta B) = \alpha \Pi(A) + \beta \Pi(B)\\
\Pi(AB) = \Pi(A) \Pi(B)\\
\Pi(A^{\star}) = [\Pi(A)]^{\dagger}
\end{eqnarray*}
The representation is cyclic, with cyclic vector $\Psi$, if the set of vectors $\{\Pi(A)\Psi; A\in{\cal A}\}$ is dense in ${\cal H}$. By the Gelfand-Naimark-Segal (GNS) construction (\cite{BRo1}, see also \cite{Wil}, 1.8.1), if $\omega$ is a state on ${\cal A}$, there exists a cyclic representation $\Pi_{\omega}({\cal A})$ of ${\cal A}$ on a Hilbert space
${\cal H}_{\omega}$, with cyclic vector $\Psi_{\omega}$, such that
$$
\omega(A) = (\Psi_{\omega}, \Pi_{\omega}(A) \Psi_{\omega})
\eqno{(2.5)}
$$
Two representations $\Pi_{1}$ and $\Pi_{2}$ of a given C* algebra ${\cal A}$ on Hilbert spaces ${\cal H}_{1}$ and
${\cal H}_{2}$ are equivalent iff there is a unitary map $U : {\cal H}_{1} \to {\cal H}_{2}$ such that
$\Pi_{2}(A) = U \Pi_{1}(A) U^{\dagger}$ for all $A \in {\cal A}$. Related concepts of quasiequivalence and disjointness are reviewed in \cite{Land}, with references, but it should be at least mentioned that two representations $\Pi_{1}$ and
$\Pi_{2}$ are \textbf{disjoint} iff no subrepresentation of $\Pi_{1}$ is unitarily equivalent to any subrepresentation of
$\Pi_{2}$. Two states which induce disjoint representations are said to be disjoint: if they are not disjoint, they are called \textbf{coherent}.

Remark: For finite dimensional matrix algebras (with trivial center) all representations are coherent.

The concept of \textbf{pure state} is fundamental: a state is \textbf{pure} iff it cannot be decomposed as a convex combination of other states, i.e., cannot be written as $\omega = \lambda \omega_{1} + (1-\lambda) \omega_{2}$ for some real
$0 < \lambda <1$ and $\omega_{i} \ne \omega, i=1,2$; otherwise, it is \textbf{mixed}. On ${\cal B}({\cal H})$ the states $\omega(A) = tr_{{\cal H}}(P_{\Psi}A) = (\Psi, A \Psi)$ with $P_{\Psi}=|\Psi)(\Psi|$ is pure ($||\Psi||=1$); the state $\omega(A) = tr_{{\cal H}}(\rho A)$ where $\rho = \rho^{\dagger}$ is a positive operator of trace one is mixed whenever
$\rho^{2} \ne \rho$. A related concept is that of a \textbf{primary} or \textbf{factor} representation $\Pi$: it is one which cannot be decomposed as direct sum of two nontrivial disjoint subrepresentations; if $\Pi_{\omega}$ is a primary representation, where $\Pi_{\omega}$ is related to $\omega$ by (2.5), $\omega$ is said to be a primary or factor state. Corresponding to the previous remark for finite dimensional matrix algebras all representations are factor representations.

The set of all pure states on ${\cal A}$ falls into equivalence classes, whereby two states $\omega_{1}$ and $\omega$ are defined to be in the same class if $\omega_{1}= \omega_{B}$ for some $B \in {\cal A}$, where
$$
\omega_{B}(A) = \frac{\omega(B^{\star} A B)}{\omega(B^{\star}B)}
$$
and $\omega(B^{\star}B) \ne 0$; these equivalence classes are called \textbf{folia} (see also \cite{Se3} and \cite{Land}): we denote them by $[\omega]$ following \cite{Land}. Each folium $[\omega]$ of the  set of pure states on ${\cal A}$ defines a \textbf{superselection sector} of the theory described by the algebra of observables ${\cal A}$. The fact that the abstract characterization of a C* algebra captures the concrete version exactly means that the representations of the quasilocal algebra ${\cal A}$ occurring in different superselection sectors can be isomorphic as C* algebras, even though they are not unitarily equivalent. This feature is a partial explanation of the naturalness of the C* algebra approach to systems with infinite number of degrees of freedom (\cite{Se3}, \cite{MWB}), and an explicit example of it will be given shortly.

Concerning the GNS state $\omega$, the following structure also arises naturally: a subalgebra of
${\cal B}({\cal H}_{\omega})$ is called a \textbf{von Neumann algebra} if it is closed in the so-called weak topology: a sequence of operators $\{A_{i}=\Pi_{\omega}(A_{i}),i=1,2,3, \cdots \}$ converges \textbf{weakly} to $A$ iff
$\lim_{i \to \infty} (\Phi, A_{i} \Psi) = (\Phi, A \Psi)$ for all $\Phi,\Psi \in {\cal H}_{\omega}$. Then
$\Pi_{\omega}({\cal A})^{'}$ - the \textbf{commutant} of $\Pi_{\omega}({\cal A})$ - the set of all bounded operators on ${\cal H}_{\omega}$ which commute with every operator in $\Pi_{\omega}({\cal A})$, defines the algebra generated by the superselection rules. The \textbf{center}
${\cal Z}(\Pi_{\omega}({\cal A})) \equiv \Pi_{\omega}({\cal A}) \cap \Pi_{\omega}({\cal A})^{'}$. If $\omega$ is a factor state, the center ${\cal Z} = \{\lambda \mathbf{1}\}$, i.e., the operators in the center are dispersion free. This is the crucial property leading to the identification of factor states with pure thermodynamic phases (see \cite{Se3}).

The following construction is very important in section 3. Let ${\cal D}$ be a set of bounded regions
$\Lambda \subset \mathbf{R}^{d}$ such that $\cup_{{\cal D}}\Lambda = \mathbf{R}^{d}$. For $\Lambda \in {\cal D}$, let
$\tilde{{\cal A}}_{\Lambda}$ be the C* algebra generated by all ${\cal A}_{\Lambda^{'}}$ with $\Lambda^{'} \in {\cal D}$ and
$\Lambda \cap \Lambda^{'} = \emptyset$ and $\Pi$ be a representation of ${\cal A}$. The following algebra corresponds to operations which may be performed outside of any bounded set:
$$
{\cal L}_{\Pi} = \cap_{{\cal D}} \Pi({\cal A}_{\Lambda})
\eqno{(2.6)}
$$
It is called the \textbf{algebra of observables at infinity} in the representation $\Pi$ \cite{LaRu}. The double commutant of $\Pi({\cal A}_{\Lambda})$ in (2.6) is, by von Neumann's bicommutant theorem (\cite{BRo1}\cite{Wil},(2.25)) equal to the weak closure of $\Pi(\tilde{{\cal A}}_{\Lambda})$. ${\cal L}_{\Pi}$ lies in the center of $\Pi({\cal A})^{'}$ and is, thus, abelian, a necessary prerequisite for a set of classical observables.

The macroscopic observables are special elements of ${\cal L}_{\Pi}$: for any sequence $\Lambda_{n} \in {\cal D}$ converging to infinity (i.e., almost all $\Lambda_{n}$ lie outside of any bounded region), let $A_{n} \in {\cal A}_{\Lambda_{n}}$ with $||A_{n}|| \le b$ uniformly in $n$, and $\Pi$ be a representation of ${\cal A}$. If
$$
w-\lim_{N \to \infty} \frac{\sum_{n=1}^{N} \Pi(A_{n})}{N} = A
\eqno{(2.7)}
$$
exists, then $A \in {\cal L}_{\Pi}$. Similarly, we have:

\textbf{Lemma 6 of \cite{He}} If $\omega_{1}$ and $\omega_{2}$ are factor states on ${\cal A}$, and
$$
\lim_{N \to \infty} \frac{\sum_{n=1}^{N} \omega_{i}(A_{n})}{N} = a_{i} \mbox{ for } i=1,2
\eqno{(2.8)}
$$
with $a_{1} \ne a_{2}$, then $\omega_{1}$ and $\omega_{2}$ are disjoint.

Indeed, let $\Pi_{\omega_{i}}$ denote the representations associated to $\omega_{i}$ by (2.5) ($i=1,2$). If $\omega_{i}$ are factor states, (2.8) implies
$$
\lim_{N \to \infty} \frac{\sum_{n=1}^{N} \Pi_{\omega_{i}}(A_{n})}{N} = a_{i} \mbox{ for } i=1,2
\eqno{(2.9)}
$$
(see \cite{He}, Lemma 5).

\subsection{Examples of non-unitarily equivalent representations and superselection sectors}

 We now give an explicit example, following \cite{NTh1}, which illustrates why (2.8) implies disjointness, and how this properties typically arises in infinite systems. For each direction $\vec{n}_{i}$, $\vec{n}_{i}^{2}=1$, there is a vector in the Hilbert space $\mathbf{C}^{2}_{i}$ such that $\vec{\sigma}_{i}$ points in this direction:
\begin{eqnarray*}
(\vec{\sigma}_{i},\vec{n}_{i})|\vec{n}_{i})_{i} = |\vec{n}_{i})_{i}
\end{eqnarray*}
$$\eqno{(2.10)}$$
For $N$ spins the $2^{N}$- dimensional Hilbert space is ${\cal H}_{N} = \otimes_{i=1}^{N} \mathbf{C}^{2}_{i}$ and in the limit $N \to \infty$ the space becomes nonseparable, but separable subspaces - the so-called IDPS, isomorphic to Fock space, see Wehrl's contribution to \cite{Wehrl} - by letting ${\cal A}$ act on a reference vector $|\Psi_{+})$: ${\cal H}_{+} = {\cal A}|\Psi_{+})$ where, for $|\Psi_{+})$ one may choose a polarized state in which all spins point in the same direction $\vec{n}$:
$$
|\Psi_{+}) = \otimes_{i=-\infty}^{\infty} |\vec{n})_{i}
\eqno{(2.11.1)}
$$
On ${\cal H}_{+}$ one obtains an irreducible representation $\Pi_{+}$ of ${\cal A}$ (see, again, \cite{Wehrl}). Weak limits such as the mean magnetization
$$
\vec{M}_{+} = w-\lim_{N \to \infty} \frac{\sum_{i=-N}^{N} \Pi_{+}(\vec{\sigma}_{i})}{2N+1} = \vec{n} \mathbf{1}
\eqno{(2.12.1)}
$$
on ${\cal H}_{+}$ depend on the representation, and, on another reference state
$$|\Psi_{-}) = \otimes_{i=-\infty}^{\infty}|\vec{m})_{i}  \eqno{(2.11.2)}$$ we obtain
$$
\vec{M}_{-} = w-\lim_{N \to \infty} \frac{\sum_{i=-N}^{N} \Pi_{-}(\vec{\sigma}_{i})}{2N+1} = \vec{m} \mathbf{1}
\eqno{(2.12.2)}
$$
The two representations cannot be unitarily equivalent, because
$U^{-1} \Pi_{+}(\vec{\sigma}) U = \Pi_{-}(\vec{\sigma}_{i})$ would imply $U^{-1} (\vec{n}\mathbf{1}) U = \vec{m} \mathbf{1}$, which is impossible because $U$ cannot change the unity $\mathbf{1}$. The $|\Psi_{\pm})$ define states
$$\omega_{\pm} (.) = (\Psi_{\pm},. \Psi_{\pm}) \eqno{(2.13.1)}$$ and the mixed state
$$
\lambda \omega_{+} + (1-\lambda) \omega_{-}
\eqno{(2.13.2)}
$$
is obtained from a vector in the orthogonal direct sum of $\Pi_{+}$ and $\Pi_{-}$
\begin{eqnarray*}
|\Omega_{S}) = \sqrt(\lambda) |\Psi_{+}) \oplus \sqrt(1-\lambda) |\Psi_{-})\\
\in {\cal H}_{+}\oplus {\cal H}_{-} \equiv {\cal H}_{S}\\
\Pi_{S} = \Pi_{+} \oplus \Pi_{-}
\end{eqnarray*}$$\eqno{(2.14.1)}$$
$$
(\Omega_{S}, \Pi_{S}(\vec{\sigma}_{i}) \Omega_{S}) = \lambda \vec{n} + (1-\lambda) \vec{m}
\eqno{(2.14.2)}
$$
The above representation is reducible, with two superselection sectors; the magnetization
\begin{eqnarray*}
\vec{M} = \lim_{N \to \infty} \frac{\sum_{i=-N}^{N}\Pi_{S}(\vec{\sigma}_{i})}{2N+1}\\
= \lambda \vec{n} + (1-\lambda) \vec{m}
\end{eqnarray*}
$$\eqno{(2.15)}$$
is in the center and not a multiple of unity. Note that $\Pi_{+}({\cal A})$ and $\Pi_{-}({\cal A})$ are isomorphic as C* algebras, with the product operation (2.1), although not unitarily equivalent as announced.

(2.14.1) generalizes to a finite number $n$ of projectors $\{\Pi_{\alpha}\}_{\alpha=1}^{n}$, with
${\cal H}_{\S}= \oplus_{\alpha=1}^{n} {\cal H}_{\alpha}$, $\Pi_{S} = \oplus_{\alpha=1}^{n} \Pi_{\alpha}$, and a state of the form
$$
\omega_{S} = (\Omega, . \Omega) = \sum_{\alpha=1}^{n} \lambda_{\alpha} \omega_{\alpha}
\eqno{(2.16.1)}
$$
with $0 \le \lambda_{\alpha} \le 1$, $\sum_{\alpha=1}^{n} \lambda_{\alpha}= 1$.

\subsection{Decoherent mixtures and macroscopically pure states. Coherence with respect to an abelian subalgebra. Automorphisms}

\textbf{Definition 1} The state $\omega_{S}$ is a \textbf{decoherent mixture} over the macroscopic (or classical ) observables iff, for $1 \le \alpha \le n$, the states $\omega_{\alpha}$ in (2.16.1) satisfy the generalization of (2.12.1,2), i.e.,
\begin{eqnarray*}
\vec{M}_{\alpha} = w-\lim_{N \to \infty} \frac{\sum_{i=-N}^{N} \Pi_{\alpha}(\vec{\sigma}_{i})}{2N+1} \\
= \vec{m}_{\alpha} \mathbf{1}
\end{eqnarray*}
$$\eqno{(2.16.2)}$$
with
$$\vec{m}_{\alpha} \ne \vec{m}_{\alpha^{'}} \mbox{ for } \alpha \ne \alpha^{'} \eqno{(2.16.3)}$$.

The states $\omega_{\alpha}$ are said to be \textbf{macroscopically pure}.\,
In fact in the thermodynamic limit the definition of "decoherent mixture " is equivalent to the fact, that the center of the representation is not trivial, whereas "macroscopically pure states" are those with trivial center. The advantage of the definition lies in the fact that it can be easily generalized to finite systems:

\textbf{Definition 1.1} The state $\omega $ over a finite-dimensional algebra ${\cal A}$ is called coherent with respect to an abelian subalgebra $\{P_{\alpha }\}, \sum _{\alpha } P_{\alpha }=1$, if there exist operators $A \in {\cal A} $ and $\alpha \neq \alpha '$ such that
$$ \omega (P_{\alpha }AP_{\alpha '})\neq 0$$
It is called macroscopically pure if there exists $\alpha $such that
$$ \omega (P_{\alpha '}AP_{\alpha '})= 0 \quad \forall \alpha ' \neq \alpha \forall A \in {\cal A}$$
Otherwise it is called decoherent.

It should be remarked that it is the abelian subalgebra of definition 1.1 which actually defines the macroscopic observables.

As a common example leading to macroscopically pure states (see also remark 6), let (2.16.1) be the decomposition into extremal invariant (ergodic) states (\cite{BRo1}, 4.2.1). The elements (2.16.2) belong to the center
${\cal Z}_{\omega_{\alpha}} = \Pi_{\omega_{\alpha}}({\cal A}) \cap \Pi_{\omega_{\alpha}}({\cal A})^{'}$ . By, e.g., \cite{Hug}, Corollary 4.1.5, if $\omega_{S}$ is assumed to be an equilibrium (KMS) state at inverse temperature $0 < \beta <\infty$ (\cite{BRo2},Chapter 5.3), so are the $\omega_{\alpha}$ and, being extremal invariant, are primary or factor states, and thus
${\cal Z}_{\omega_{\alpha}} = \{\lambda \mathbf{1}\}$, and therefore (2.16.2) holds. By (2.16.3) and the previously mentioned lemma 6 of \cite{He}, the $\omega_{\alpha}$ are mutually disjoint, and the structure of (2.16.1) is complete.

There are, of course, obvious generalizations to particle systems, with the observables $\vec{M}_{\alpha}$ replaced by the particle, energy or momentum density.

A basic dynamical concept is that of \textbf{automorphism}: an automorphism $\tau \in Aut({\cal A})$ is a one to one mapping of ${\cal A}$ onto ${\cal A}$ which preserves the algebraic structure. The following lemma is important (for an alternative proof to Hepp's, see \cite{BRo1}, Prop. 2.4.27):

\textbf{Lemma 2 of \cite{He}} If $\omega_{1}$ and $\omega_{2}$ are disjoint states on ${\cal A}$, and
$\tau \in Aut({\cal A})$, then $\omega_{1} \circ \tau$ and $\omega_{2} \circ \tau$ are disjoint.

Above, the circle denotes composition, i.e., $(\omega \circ \tau)(A) = \omega(\tau(A))$. By this lemma, coherence cannot be destroyed during the measurement process, as long as the time evolution is automorphic. It can only be shifted.

What are the main properties of the group of time-translation automorphisms
$t \in \mathbf{R} \to \tau_{t} \in Aut({\cal A})$? For a large class of quantum spin systems - those with potentials of finite range - the local Hamiltonian $H(\Lambda)$ generates an automorphism of ${\cal A}(\Lambda)$ by
$$
A \to A_{t} = \exp(it H(\Lambda)) A \exp(-it H(\Lambda))
\eqno{(2.17)}
$$
and the limit
$$
\tau_{t}(A) = \lim_{\Lambda \to \infty} \exp(it H(\Lambda)) A \exp(-it H(\Lambda))=\lim_{\Lambda \to \infty}A_{t}^{\Lambda}
\eqno{(2.18)}
$$
exists on the dense local subalgebra (2.2.1). In the above limit $\Lambda \to \infty$ could be taken to mean along a sequence of hypercubes whose sides tend to infinity, but many other choices leading to the same element of ${\cal A}$ are possible. The limit is uniform for t in some circle in the complex plane around zero. One then extends $\tau_{t}$ to the whole quasi-local algebra ${\cal A}$ by the density property (2.2.2), and then, by the group property of $\tau_{t}$ on ${\cal A}_{L}$, to all
$t \in \mathbf{R}$: see (\cite{Stre},\cite{Rob}).

$A_{t}^{\Lambda}$ in (2.18) is constructed as
$$
A_{t}^{\Lambda} = A + it[H(\Lambda),A] + (it)^{2}/2 [H(\Lambda),[H(\Lambda),A]] + \cdots
\eqno{(2.19)}
$$
Each multiple r-commutator in (2.19) takes an initial $A \in {\cal A}(\Lambda_{0})$ roughly to an

$A \in {\cal A}(\Lambda_{0}+rd)$, where $d$ is the range of the interaction potential (see also \cite{LiR} and \cite{NS} for localization properties of $\tau_{t}(A)$) ; when $r \to \infty$, $A$ becomes infinitely extended, and, for that reason $\tau_{t}$ becomes an automorphism \textbf{only} of the quasilocal algebra ${\cal A}$ given by (2.2.2) and not of \textbf{any} local algebra ${\cal A}(\Lambda)$ - this happens as long as there is any interaction between the spins, i.e., for any nonzero range $d$. This is a basic property of the quantum evolution, which will play an important role in section 3.1.

\section{Reduction of the wave-packet: the approaches of Hepp and Sewell}

In this section we review the approaches of Hepp \cite{He} and Sewell (\cite{Sew1}\cite{Sew2}) to the quantum measurement problem.

\subsection{Hepp's approach: disjointness and vanishing of the overlaps}

Hepp's lemma 2 (\cite{He} and section 2) shows that coherence cannot be destroyed by an automorphic time evolution during the process of measurement, but his lemma 6 (\cite{He} and section 2) and the example in section 2 suggest that it might be possible to find sequences $\omega_{1,n}$ and $\omega_{2,n}$ of coherent states which converge weakly
(denoted $w-\lim \omega_{i,n}= \omega_{i}$) towards disjoint states $\omega_{1}$, $\omega_{2}$, and one has the important

\textbf{Lemma 3 of \cite{He}} Let $\Pi_{n}$ be representations of ${\cal A}$ and
$\Psi_{i,n} \in {\cal H}_{\Pi_{n}}$ with $\omega_{i,n}= \omega(\Psi_{i,n}) \circ \Pi_{n}$, $i=1,2$, i.e.,
$\omega_{i,n}(A) = (\Psi_{i,n}, \Pi_{n}(A) \Psi_{i,n})$, $i=1,2$
Consider sequences
$$
w-\lim \omega_{i,n} = \omega_{i} \mbox{ for } i=1,2
\eqno{(3.1.1)}
$$
with $\omega_{1}$, $\omega_{2}$ disjoint. Then, for all $A \in {\cal A}$,
$$
\lim_{n \to \infty} (\Psi_{1,n}, \Pi_{n}(A) \Psi_{2,n}) = 0
\eqno{(3.1.2)}
$$

Hepp does not prove the main inequality (\cite{He}, (3.5)) used to prove the above lemma, nor gives any reference. Actually the first inequality in (\cite{He}, (3.5)) may be directly shown to be an equality, a fact which renders his sketch of proof less transparent. For the convenience of the more mathematically minded reader, and since this section has a review character, we provide a complete proof below (readers uninterested in mathematical details should skip it):

\textbf{Proof} If $\omega_{2}(A^{\star}A)=0$, $|(\Psi_{1,n}, \Pi_{n}(A) \Psi_{2,n})|^{2} \le \omega_{2,n}(A^{\star}A) \to 0$; otherwise, $\omega_{2,n}(A^{\star}A) \ne 0$ for $n$ sufficiently large and
$$
\omega_{2,n}^{A} \equiv (\Psi_{2,n}, \Pi_{n}(A)^{\dagger} (.) \Pi_{n}(A))/(\Psi_{2,n}, \Pi_{n}(A^{\star}A) \Psi_{2,n})
$$
is such that
$$
w-\lim \omega_{2,n}^{A} = \omega_{2,A}
$$
where $\omega_{2,A} \in [\omega_{2}]$, the folium of $\omega_{2}$. Since $\omega_{1}$ and $\omega_{2}$ are disjoint,
$\omega_{1}$ and $\omega_{2,A}$ are disjoint, and thus
$$
||\omega_{1}-\omega_{2,A}|| = 2
\eqno{(3.1.3)}
$$
where $||\omega|| = \sup_{||A||=1}|\omega(A)|$, by the theorem of Glimm and Kadison (\cite{GlKa}, see also \cite{Wil}, theorem 2.6.6 for a simple proof). By (\cite{Scha}, theorem 2, pg. 47),
$$
||\omega_{1,n} - \omega_{2,n}^{A}||^{2} = [\tau(T)]^{2}
\eqno{(3.1.4)}
$$
where $\tau(T)$ denotes the trace norm of the operator
$T \equiv \Psi_{1,n} \otimes \overline{\Psi_{1,n}} - \Psi_{2,n}^{A} \otimes \overline{\Psi_{2,n}^{A}}$, where
$\Psi_{2,n}^{A} \equiv \frac{\Pi_{n}(A) \Psi_{2,n}}{||\Pi_{n}(A) \Psi_{2,n}||}$, and
\begin{eqnarray*}
(x \otimes \overline{x})f \equiv (f,x) x\\
\mbox{ for all } f \in {\cal H}_{\Pi_{n}}
\end{eqnarray*}
Note that $||\Pi_{n}(A) \Psi_{2,n}|| \ne 0$ for sufficiently large $n$. The above trace norm may be evaluated by converting $(\Psi_{1,n}, \Psi_{2,n}^{A})$ to an orthonormal system by the Gram-Schmidt procedure and equals
$$
\tau(T) = 2 \sqrt{1 - |(\Psi_{1,n},\Psi_{2,n})|^{2}}
\eqno{(3.1.5)}
$$
(3.1.2) follows, then, from (3.1.3), (3.1.4) and (3.1.5) together with the weak continuity of the norm. q.e.d.

Hepp constructs examples of automorphic evolutions $\tau_{t} \in Aut({\cal A})$ such that, for $i=1,2$,
$ w\lim_{t \to \infty} \omega_{i} \circ \tau_{t} = \bar{\omega_{i}}$ and $\omega_{1}$, $\omega_{2}$ are coherent, but
$\bar{\omega_{1}}$ and $\bar{\omega_{2}}$ are disjoint. Disjointness is brought about by the existence of a macroscopic observable with different expectation values in $\omega_{1}$, $\omega_{2}$ (lemma 6 of \cite{He} and section 2). 

\subsection{The Coleman-Hepp model} 

We now come to an example, which was called by Bell \cite{B} the Coleman-Hepp model, a terminology we also adopt. We initially follow Bell \cite{B} for clarity of exposition in the description of the model.

The apparatus is a semi-infinite linear array of spin one-half particles, fixed at positions $n=1,2,\cdots 2L+1$; the system is a moving spin one-half particle with position coordinate $x$ and spin operators
$\vec{\sigma_{0}} \equiv (\sigma_{0}^{1}, \sigma_{0}^{2}, \sigma_{0}^{3})$- the third component $\sigma_{0}^{3}$ is to be ''measured''. The combined system is described by a wave function
$$
\Psi(t,x,\sigma_{0},\sigma_{1}, \cdots )
\eqno{(3.2)}
$$
in a representation where all $\sigma_{n}^{3}$ are diagonal, and $\sigma_{i}=\pm 1$ for all $i=0,1,2, \cdots$. The Hamiltonian is
$$
H = -i\frac{\partial}{\partial x} + \sum_{n=1}^{\infty} V(x-n) \sigma_{n}^{1}(1/2 - 1/2 \sigma_{0}^{3})
\eqno{(3.3)}
$$
The interaction is assumed to have compact support:
$$
V(x) = 0 \mbox{ for } |x| > r
\eqno{(3.4.1)}
$$
and to satisfy
$$
J =\frac{\int_{-\infty}^{\infty} dx V(x)}{\pi} = 1/2
\eqno{(3.4.2)}
$$
The Schr\"{o}dinger equation $i \frac{\partial}{\partial t} \Psi = H \Psi $ is solved by
\begin{eqnarray*}
\Psi(t,x,\sigma_{0},\sigma_{1}, \cdots ) = \\
\prod_{n=1}^{\infty} \exp[-i F(x-n) \sigma_{n}^{1}(1/2 - 1/2 \sigma_{0}^{3})]\Phi(x-t, \sigma_{0},\sigma_{1}, \cdots )
\end{eqnarray*}
$$\eqno{(3.4.3)}$$
where $\Phi$ is arbitrary and
$$
F(x) \equiv \int_{-\infty}^{x} dy V(y)
\eqno{(3.4.4)}
$$
The function $F$ has the properties
\begin{eqnarray*}
F(x) = \left\{
\begin{array}{ll}
0      &  \mbox{ if } x<-r\,,\\
\pi/2  &  \mbox{ if } x>r\,,
\end{array}
\right.
\end{eqnarray*}$$\eqno{(3.5)}$$
Consider in particular states in which the lattice spins are all up and the moving spin is either up or down:
\begin{eqnarray*}
\Psi_{+}(t,x, \cdots ) = \chi(x-t) \Psi_{+}(\sigma_{0}) \prod_{n=1}^{\infty} \Psi_{+}(\sigma_{n})\\
\Psi_{-}(t,x, \cdots ) = \chi(x-t) \Psi_{-}(\sigma_{0}) \prod_{n=1}^{\infty} \Psi_{+}^{'}(\sigma_{n},x-n)
\end{eqnarray*}
$$\eqno{(3.6)}$$
where
\begin{eqnarray*}
\Psi_{\pm}(\sigma) = \delta_{\sigma \mp 1}\\
\Psi_{+}^{'} (\sigma_{n},x-n) = \exp[- i F(x-n) \sigma_{n}^{1}] \Psi_{+}(\sigma_{n})
\end{eqnarray*}
$$\eqno{(3.7)}$$
Due to (3.5),
\begin{eqnarray*}
\Psi_{+}^{'}(\sigma_{n},x-n) = \left\{
\begin{array}{ll}
\Psi_{+}(\sigma_{n})    &   \mbox{ for } x-n<-r\,,\\
-i \Psi_{-}(\sigma_{n}) &   \mbox{ for } x-n>r\,,
\end{array}
\right.
\end{eqnarray*}
$$\eqno{(3.8)}$$
Let us also suppose that the wave packet $\chi$ has compact support:
$$
\chi(x) = 0 \mbox{ for } |x|>w
\eqno{(3.9)}
$$
Then, from (3.8), we may use in (3.6)
$$
\Psi_{+}^{'}(\sigma_{n},x-n) = \Psi_{+}(\sigma_{n})
\eqno{(3.10.1)}
$$
$$
\mbox{ for } n> x+r \ge t-w+r
\eqno{(3.10.2)}
$$
together with
$$
\Psi_{+}^{'}(\sigma_{n},x-n) = -i \Psi_{-}(\sigma_{n})
\eqno{(3.10.3)}
$$
$$
\mbox{ for } n< x-r \le t+w-r
\eqno{(3.10.4)}
$$
The Coleman-Hepp model, described above, is a caricature of an electron in one-dimensional motion, whose spin is measured by result of a local interaction with an infinite spin array. The latter is brought about  by a potential of compact support
 but the coordinate $x$ plays only an intermediate role. Under conditions (3.10), the evolution does not depend on $x$, except by the wave function $\chi$ in (3.9). Due to the linearity of (3.3) in the momentum operator, the evolution of $\chi$ is only a translation.

Let the quasilocal algebra $\tilde{{\cal A}}$ be chosen as the tensor product
$$
\tilde{{\cal A}} = {\cal B}(\mathbf{C}_{0}^{2}) \otimes {\cal A}
\eqno{(3.11)}
$$
where ${\cal A}$ is the quasilocal spin algebra (2.2.1), (2.2.2). We assume that the coordinate $x$ is not ''measured'', i.e., the corresponding part of the algebra is the identity. Correspondingly, we omit the factor $\chi(x-t)$ in the forecoming formulas, because $(\chi, \chi) = 1$, which is not affected by a translation by $t$.

Let $\Psi_{+}$ and $\Psi_{-}$  be given by (2.11.1), (2.11.2), where $\vec{n} \equiv (0,0,1)$, and $\vec{m} \equiv (0,0,-1)$, respectively, and $i = 1,2, \cdots $ in definitions (2.11.1), (2.11.2). The precise mathematical meaning of (3.6) and (3.10) is that
$\Psi(t,x, \cdots)$ equals $\Psi_{+}$ and that $\Psi(t,x, \cdots)$ equals $\Psi_{-}$, the latter if the conditions in (3.10) are verified. More precisely, if $\Pi_{n}$ is the representation of ${\cal A}$ on the Hilbert space
${\cal H}_{n} = \mathbf{C}_{1}^{2} \otimes \cdots \otimes \mathbf{C}_{n}^{2}$, the states
$$
\omega_{\pm,n} \circ \Pi_{n}
\eqno{(3.12.1)}
$$
where
$$
\Psi_{\pm,n} = \Psi_{0,\pm} \otimes_{i=1}^{n} \Psi_{i,\pm}
\eqno{(3.12.2)}
$$
with $\sigma_{0}^{3}\Psi_{0,\pm} = \pm \Psi_{0,\pm}$, tend weakly, as $n \to \infty$, on the algebra $\tilde{{\cal A}}$, to states $\omega_{\pm}$ given by (2.13.1); by (2.11.1,2) (with index sum running from zero to $n$ instead of $-N$ to $N$). By lemma 6 of \cite{He} (see also section 2), they are disjoint. Consider now the initial vector
$$
\Psi_{0,n} \equiv \Psi_{0} \otimes_{i=1}^{n} \Psi_{i,+}
\eqno{(3.13.1)}
$$
where
$$
\Psi_{0} \equiv c_{+} \Psi_{0,+} + c_{-} \Psi_{0,-}
\eqno{(3.13.2)}
$$
Its evolution is given, by (3.6) and (3.10) (under condition (3.10)):
$$
\Psi_{0,n}(t) = c_{+} \Psi_{0,+} \otimes_{i=1}^{n} \Psi_{i,+} + c_{-} \Psi_{0,-} \otimes_{i=1}^{n} \Psi_{i,-}
\eqno{(3.13.3)}
$$
By lemma 3 of \cite{He} the corresponding sequence of states $\omega_{\Psi_{0,n}(t)}$ tends weakly, as $n \to \infty$, with
$t$ satisfying (3.10.2), which implies $t \to \infty$), to the mixed state
$$
\omega = |c_{+}|^{2} \omega_{+} + |c_{-}|^{2} \omega_{-}
\eqno{(3.13.4)}
$$
i.e., the state (2.13.2), with $\lambda = |c_{+}|^{2}$ and $1-\lambda = |c_{-}|^{2}$.

The Coleman-Hepp model has two main shortcomings from a physical standpoint: the linear, instead of quadratic, momentum term, i.e., absence of dispersion, and the lack of interaction between the spins in (3.3). We shall now dwell on the consequences of this last defect, but remark that, in spite of that, it is quite ingenious, allowing for detailed estimates on the time evolution, an extremely rare feature. For other examples, this time with interaction, see \cite{NTh3}, \cite{NTh4}, as well as the model of Curie-Weiss spins weakly interacting with a phonon bath introduced in \cite{ABN2} (see also \cite{ABN1}).

Consider, now, instead of the quasilocal algebra, the \textbf{local} subalgebra
$$
\tilde{{\cal A}}_{M} = {\cal B}(\mathbf{C}_{0}^{2}) \otimes {\cal A}_{M}
\eqno{(3.14.1)}
$$
with
$$
{\cal A}_{M}= {\cal A}(\Lambda_{M})
\eqno{(3.14.2)}
$$
and ${\cal A}(\Lambda)$ is the local spin algebra of section 2, with
$$
\Lambda_{M} \equiv \{1,2, \cdots M\}
\eqno{(3.14.3)}
$$
We are allowed to do that because the local subalgebras (3.14.1,2) are left invariant by the dynamics, i.e., the time evolution given by (3.3) defines an automorphism of these local algebras, not just the quasilocal algebra (2.2.2). This is due to the previously mentioned fact that in (3.3) there are no interactions between the spins, i.e., the interaction range $d$ is zero see (2.19) and remarks thereafter. By (3.10),if
$$
t \ge M+1-w+r
\eqno{(3.15)}
$$
we have (3.1.2) in the strong form
\begin{eqnarray*}
(\Psi_{\pm,n}(t), \Pi_{n}(A) \Psi_{\mp,n}(t)) = 0 \mbox{ if } n>M \\
\mbox{ and } t \mbox{ satisfies (3.15) } \mbox{ and } A \in \tilde{{\cal A}}_{M}
\end{eqnarray*}
$$\eqno{(3.16.1)}$$
where the embedding of $A$ in ${\cal B}({\cal H}_{n})$ is given by $A \to A \otimes \mathbf{1} \otimes \mathbf{1} \cdots$. Since (3.16.1) occurs for finite times $t$ (satisfying (3.15)), neither irreversibility nor disjoint states - now of the form (3.12.1), (3.12.2), with $n>M$ and $t$ satisfying (3.15) - occur in this case!

The estimate for the critical ''decoherence time'' $t_{0}$ resulting from (3.15),
$$
t_{0} = M+1-w+r
\eqno{(3.15.1)}
$$
is microscopic only for suitable $M$, e.g., for $M=10^{9}$, corresponding to a length of the chain equal to
$Ma \approx 10^{9} \times 10^{-8} \approx 10 cm$, taking for the spacing between the atoms in the chain
$a \approx 10^{-8} cm$, $t_{0} \approx M\frac{m a^{2}}{\hbar}$ yields, taking for $m$ the electron mass, ,
$t_{0}^{est} \approx 10^{-7}$ seconds, roughly comparable with spin decoherence times in quantum wells. Indeed, some of the largest decoherence times are found in electron spin coherence in GaAs quantum dots \cite{BuLdV}, which is mostly due to hyperfine interactions with nuclear spins, and range from 0.1-100 microseconds, i.e, ranging from $(1-10^{2})t_{0}^{est}$. Notice that this decoherence time is nothing but the time taken by the electron to traverse the chain, as observed by Sewell \cite{Sew1}, \cite{Sew2}, because it equals the distance $Ma$ divided by the ''electron velocity''
$\hbar k/m$, where the wave vector $k \approx 1/a$. For larger $M$ this estimate quickly becomes highly unrealistic, suggesting us to adopt a modified definition, to which we now turn.

\subsection{First definition of reduction of the wave packet and decoherence}

\textbf{Definition 2}\, We say that a sequence of states $\omega_{\Psi_{0,n}(t)} \circ \Pi_{n}$ given by (3.13.3) exhibits \textbf{reduction of the wave packet} or \textbf{decoherence} with respect to a subalgebra $\cal A_M \subset \cal A $ stable under the time evolution iff
$$
\lim_{n \to \infty,t>t_{0n}} (\Psi_{1,n}(t), \Pi_{n}(A) \Psi_{2,n}(t)) = 0
\eqno{(3.16)}
$$
for all $A \in {\cal A_M}$ , where $t_{0n}$ is a ''decoherence time''.

We have included in the definition the possibility to consider the decoherence time only with respect to a subalgebra. This opens the possibility to consider a net of increasing subalgebras with finite though increasing decoherence time. The physically most interesting case in which $t_{0n}$ is independent of $n$ is, of course, included.

The above definition is, in fact, very natural: it states that there is no ''experiment'' (represented by a quasilocal operator) connecting the ''pointer positions'' represented by the vectors $\Psi_{1,n}(t)$ and $\Psi_{2,n}(t)$ for sufficiently large systems, and times greater than a critical value. Lemma 3 of \cite{He} proves that disjointness of the corresponding states $w\lim_{n,t \to \infty} \omega_{1,2}^{t,n} \equiv \omega_{\Psi_{1,2,n}(t)} \circ \Pi_{n} $ is a way of achieving this aim, but only with $t_{0n} \to \infty$ as $n \to \infty$. The latter happens in the Coleman-Hepp model, but definition 2 leaves open the possibility of a fixed $t_{0}$ independent of $n$ of a realistic order of magnitude. An application of definition  2 to the Coleman-Hepp model is made in proposition 1 of the forthcoming section. There are, however, no examples yet with $t_{0}$ independent of $n$ which display reduction of the wave-packet in accordance with definition 2.

\subsection{Bell's criticism of Hepp's approach on the basis of the Coleman-Hepp model}

We conclude our review of Hepp's approach with some remarks on Bell's criticism \cite{B} of Hepp's work. From (3.10.2) and (3.10.4) (we take
$w=r$ for simplicity: this does not affect the argument), the spin flips occur just after the times $t_{n}=n$. Again for simplicity, we consider only discrete times $t_{n}=n$. Define the operator
$$
Z_{n} = \sigma_{0}^{1} \prod_{k=1}^{n} \sigma_{k}^{1}
\eqno{(3.17.1)}
$$
This operator has the property that
$$
\lambda_{n} \equiv (\Psi_{1,n}(t=n), Z_{n} \Psi_{2,n}(t=n)) = c > 0
\eqno{(3.17.2)}
$$
where $c$ is a constant independent of $n$ (it ''undoes'' the measurement, see \cite{LMi} and \cite{Land}). Although
$Z_{n} = \Pi_{n}(A_{n})$ for some $A_{n} \in {\cal A}$ given by (2.2.1),(2.2.2) (in fact, for an \textbf{explicit} element
$A_{n} \in {\cal A}_{L}$), for each finite $n$, there is \textbf{no} element $A \in {\cal A}$ such that the \textbf{infinite} sequence
$\{\lambda_{n}\}$, given by (3.17.2), satisfies
$$
\lim_{n \to \infty} (\Psi_{1,n}(t=n), \Pi_{n}(A) \Psi_{2,n}(t=n)) = \lim_{n \to \infty} \lambda_{n} = c
\eqno{(3.17.3)}
$$
Indeed, (3.17.3) would contradict lemmas 3 and 6 of \cite{He} (see also section 2), but this may be also shown in an elementary way as follows. (2.1.1,2.1.2) imply that, for any $A \in {\cal A}$, given any $\epsilon > 0$, there exists
$A_{L} \in {\cal A}_{L}$ such that $||A-A_{L}|| < \epsilon $; writing
\begin{eqnarray*}
(\Psi_{1,n}(t=n), \Pi_{n}(A) \Psi_{2,n}(t=n)) = (\Psi_{1,n}(t=n), \Pi_{n}(A-A_{L}) \Psi_{2,n}(t=n)) +\\
+(\Psi_{1,n}(t=n), \Pi_{n}(A_{L}) \Psi_{2,n}(t=n))
\end{eqnarray*}
and using $||\Pi_{n}(A-A_{L})|| \le ||A- A_{L}||$ together with the fact that
$$
(\Psi_{1,n}(t=n), \Pi_{n}(A_{L}) \Psi_{2,n}(t=n)) = 0
$$
for $n > \mbox{ diameter of } A_{L}$, because of the orthogonality of $\Psi_{1,n}(t=n)$ and $\Psi_{2,n}(t=n)$ for each finite $n$, we obtain a contradiction with (3.17.3).

One may ask why not restrict oneself to fixed, finite $n$. This is indeed possible if one adopts the forthcoming definition 3 of the reduction, which refers only to the microsystem and not to the measuring instrument. If, however, one wishes to probe into the observables of both system and apparatus, this is a bad choice, as we now show.

Consider, now, that the dynamics, instead of being defined by the Hamiltonian $H$ defined in (3.3), is described by the slightly (for $\alpha$ small) perturbed Hamiltonian
$$
H_{\alpha,N} \equiv H_{N} + \alpha \sum_{m=1}^{N}\sigma_{m}^{1}\sigma_{m+1}^{1}
\eqno{(3.18)}
$$
with $\alpha$ real. Above, $H_{N}$ is the finite chain version of the Coleman-Hepp model (the sum in (3.3) running from
$1$ to $N$, and the limit, as $N \to \infty$, is understood in the automorphism sense (see (2.19) et seq.).  The effect of the interaction in (3.18) is to render the observable algebra necessarily infinitely extended, as described in (2.19) et. seq., so that \textbf{no} strictly local algebra is preserved by the dynamics: we are forced to consider the algebras  $\tilde{{\cal A}}_{M}$ with $M \to \infty$ as discussed after (3.14.1). Although we are not able to solve the quantum evolution for any $\alpha \ne 0$, the possibility of such small interactions between the spins should be allowed for in any model, and the \textbf{choice} of a local algebra is not stable under these.

Definition 2 probes \textbf{a priori} the macroscopic nature of the pointers: the vanishing of the cross terms (3.16) in the limit of infinite degrees of freedom is natural for ''essentially local''- supposedly physical - observables, which are blind to changes in the global structure. One may ask what is so important about requiring (3.17.3), i.e., $A$ quasilocal. This is, of course, a philosophical question related to the proposal by Haag and Kastler \cite{HK} of what is an adequate choice of observables. But, in our opinion, it is precisely in the present context that their argument is most convincing: (3.16) probes, as $n \to \infty$, a \textbf{global} change, but this limit does not exist mathematically: it is the would-be operator which intertwines two disjoint representations. It seems also reasonable to suppose that the associated sequence of observables is devoid of any physical meaning.

It is possible to study reduction of the wave-packet using a definition (the forthcoming definition 3) introduced by von Neumann \cite{vN} and used by Sewell in \cite{Sew1},\cite{Sew2}, which involves only observables of the system, not the apparatus. This definition, by its own scope, does not by itself \textbf{require} macroscopic pointers (although this may be, and usually is, included in the definition as extra requirement). It is, therefore, more general and possesses a higher degree of flexibility than definition 2. This is demonstrated by the case $\beta = \infty$ in proposition 1. Following this idea, we use a modification of definition 2 (definition 5 in section 5.12), which  is actually implicitly widely used in the theoretical physics literature, to study a class of models of decoherence.

\subsection{Sewell's approach and von Neumann's definition of a macrostate}

We now turn to Sewell's approach (\cite{Sew1}\cite{Sew2}), whose objective was, as Hepp's, to reconsider the quantum measurement problem within the framework of Schr\"{o}dinger dynamics, but utilizing the language of quantum probability \cite{vN}. He considers a composite system $S_{c}$, consisting of a microsystem $S$ coupled to a measuring instrument ${\cal I}$,
$S_{c}= S + {\cal I}$, symbolically, where ${\cal I}$ is a large, but finite, $N$- particle system. Also essential to his approach was to take into explicit account the macroscopic nature of the observables $M$, which are taken to comprise a set of coarse-grained intercommuting extensive variables, whose simultaneous eigenspaces correspond to the pointer positions of ${\cal I}$. The approach leans on von Neumann's picture of the measurement process discussed in the introduction, but more specifically according to which the coupling of $S$ to ${\cal I}$ leads to the following two effects:
(I) It converts a pure state of $S$, as given by a linear combination $\sum_{r=1}^{n} c_{r}u_{r}$ of its orthonormal energy eigenstates $u_{r}$ into a statistical mixture of these states for which $|c_{r}|^{2}$ is the probability of finding this system in the state $u_{r}$;\\
(II) It sends a certain set of classical, i.e., intercommuting macroscopic variables $M$ of ${\cal I}$ to values, indicated by pointers, that specify which of the states $u_{r}$ is realized.
Sewell assumes that the algebras ${\cal A}$ of bounded operators of the microsystem, ${\cal B}$ of the instrument ${\cal I}$, and their composite $S_{c} = (S+{\cal I})$ are those of the bounded operators on separable Hilbert spaces ${\cal H}$,
${\cal K}$ and ${\cal H} \otimes {\cal K}$, respectively. Correspondingly, the states of these systems are represented by the density matrices in the respective spaces. The density matrices for the pure states of any of these systems are then the projection operators $P(f)$ of their normalized vectors $f$. For simplicity, it was assumed that ${\cal H}$ is of finite dimensionality $n$.

The macroscopic description of ${\cal I}$ pertinent to the measuring process was then based on an abelian subalgebra
${\cal M}$ of ${\cal B}$, which is generated by the aforementioned coarse-grained macroscopic observables, typically extensive variables of parts or the whole of ${\cal I}$. The choice of ${\cal M}$ yields a partition of ${\cal K}$ into the simultaneous eigenspaces ${\cal K}_{\alpha}$ of its elements. These subspaces of ${\cal K}$ were termed ''phase cells'', being the canonical analogues of classical phase ''pointer positions'' (compare (1c) et seq.) of this instrument.

An important element of \cite{Sew1}, \cite{Sew2} is an amplification property of the $S-{\cal I}$ - coupling, whereby different microstates of $S$ give rise to macroscopically different states of ${\cal I}$. Since ${\cal I}$ is designed so that the pointer readings are in one-to-one correspondence with the eigenstates $u_{1}, \cdots u_{n}$ of $S$, we assume that the index $\alpha$ of its macrostates also runs from $1$ to $n$. hence, denoting the projection operator for ${\cal K}_{\alpha}$ by $\Pi_{\alpha}$, it follows from the above specifications that:
$$
\Pi_{\alpha} \Pi_{\beta} = \Pi_{\alpha} \delta_{\alpha \beta}
\eqno{(3.19.1)}
$$
$$
\sum_{\alpha=1}^{n} \Pi_{\alpha} = \mathbf{1}_{{\cal K}}
\eqno{(3.19.2)}
$$
and that each element $M$ of ${\cal M}$ takes the form
$$
M = \sum_{\alpha=1}^{n} M_{\alpha} \Pi_{\alpha}
\eqno{(3.19.3)}
$$
where the $M_{\alpha}$ are scalars.

The above description corresponds to von Neumann's description of a macrostate of a macroscopic quantum system \cite{vN}, as sketched in the introduction - ''rounding off'' and ''tolerance'' effects may be added without problems, but we shall omit them from our discussion. It will be assumed that $S_{c}$ is a conservative system described by a Hamiltonian
$$
H_{c} = H \otimes \mathbf{1}_{{\cal K}} + \mathbf{1}_{{\cal H}} \otimes R + V
\eqno{(3.20)}
$$
and such that the interaction potential $V$ induces no transitions between the eigenstates $u_{1}, \cdots u_{n}$  of the Hamiltonian $H$ of $S$ - ${\cal I}$ is, thus, an ''instrument of the first kind'' \cite{J1}. The systems $S$ and ${\cal I}$ are coupled at $t=0$ following independent preparation of $S$ in a pure state and ${\cal I}$ in a (in general) mixed one, as represented by a normalized vector $\Psi_{0}$ and a density matrix $\Omega$, respectively ; the initial state of the composite $S_{c}$ is thus given by the density matrix
$$
\rho(0) = P(\Psi_{0}) \otimes \Omega
\eqno{(3.21.1)}
$$
and its evolute at time $t > 0$ by
$$
U^{\dag}_{c}(t) \rho(0) U_{c}(t) \equiv \rho(t)
\eqno{(3.21.2)}
$$
where
$$
U_{c}(t) \equiv \exp(i H_{c} t)
\eqno{(3.21.3)}
$$
Further, since $\Psi_{0}$ is a normalized vector, it is a linear combination of the basis vectors
$$
\Psi_{0} = \sum_{r=1}^{n} c_{r} u_{r}
\eqno{(3.21.4)}
$$
with
$$
\sum_{r=1}^{n} |c_{r}|^{2} = 1
\eqno{(3.21.5)}
$$
The expectation values for the time-dependent state $\rho(t)$ are defined, as usual, by
$$
E(A \otimes M) = Tr(\rho(t) A \otimes M) \mbox{ for all } A \in {\cal A} \mbox{ and for all } M \in {\cal M}
\eqno{(3.22.1)}
$$
where $Tr$ is the trace on ${\cal H} \otimes {\cal K}$; the conditional expectation value $E(A|{\cal K}_{\alpha})$ of $A$ given the macrostate ${\cal K}_{\alpha}$ of ${\cal I}$ may then be defined by
$$
E(A|{\cal K}_{\alpha}) \equiv E(A \otimes \Pi_{\alpha})/w_{\alpha}
\eqno{(3.22.2)}
$$
for all $A \in {\cal A}$, $w_{\alpha} \ne 0$, where
$$
w_{\alpha} = E(\mathbf{1}_{{\cal H}} \otimes \Pi_{\alpha})
\eqno{(3.23)}
$$

\subsection{Second definition of reduction of the wave packet and decoherence}

It follows from the above that the conditions for the realization of conditions (I) and (II) may be summarized in the following

\textbf{Definition 3} The system displays reduction of the wave packet or decoherence iff
$$
E(A) = \sum_{r=1}^{n} |c_{r}|^{2}(u_{r}, A u_{r}) \mbox{ for all } A \in {\cal A}
\eqno{(3.24.1)}
$$
and
\begin{eqnarray*}
\mbox{ there is a unique invertible transformation } \phi \mbox{ of the set } 1,2, \cdots n\\
\mbox{ such that } \\
E(A|{\cal K}_{\alpha}) = (u_{\phi(\alpha)}, A u_{\phi(\alpha)}) \mbox{ for all } A \in {\cal A}
\end{eqnarray*}$$\eqno{(3.24.2)}$$
for times $t$ such that $t > t_{0,n}$, where $t_{0,n}$ is a critical time (''decoherence time'').

In other words, the pointer reading $\alpha$ signifies that the resulting state of $S$ is $u_{\phi(\alpha)}$.

It was proved in \cite{Sew1}\cite{Sew2} that, under suitable conditions on $H_{c}$, given by (3.20), and $\Omega$ (given by (3.21.1)), (3.24) holds, and, furthermore, that these conditions are fulfilled by the finite version of the Coleman-Hepp model (3.3), where the infinite sum is replaced by a finite one running from $1$ to $N$, i.e., $N$ is the number of points in the chain, with a suitable choice of ''phase cells'', to which we come back in 5.1. Sewell also showed that that the conditions are satisfied by the previously mentioned model \cite{ABN2}, see also \cite{ABN1}. Actually, (3.24) was proven up to exponentially small corrections in $N$.  The estimate for the critical time $t_{0}$ agrees with (3.15), with the correspondence $N = M$.

Sewell's method entails several important aspects which are absent in Hepp's discussion: explicit use of the ''cell' decomposition (3.19) (which will play a central role in our illustration of the second law in section 4), and a property of stability against local perturbations of the state $\Omega$. This is an essential property of any measuring apparatus, which leads to consideration of mixed states, in addition to the ground state, a fact which will also play a very important role in sections 5.3 and 5.4. Finally, the method used to control the above mentioned exponentially small corrections allows a discussion of the ''bad microstates'', see remark 3 in section 5.4. Although the application of this method to the Coleman-Hepp model is elementary, \cite{Sew2} indicates a more general result,  which shows that these corrections are governed by the large deviation principle (see, e.g., \cite{El}).

\section{Irreversibility and a quantum version of the second law of thermodynamics}

In \cite{He}, pg. 247, Hepp remarks: ''The solution of the problem of measurement is closely connected with the yet unknown correct description of irreversibility in quantum mechanics''. And, further, at the end: ''...while the automorphic evolution between finite times is reversible, it is precisely the irreversibility in the limit of infinite times which reduces the wave packets''.

To the extent that irreversibility (in closed systems) is observed in Nature at definite (finite) times, ''here and now'', so to speak, and not just as an approximation, we believe that the weak limit $t \to \infty$ in Hepp's article is not adequate to relate the problem of measurement to irreversibility (this concern was, however, beyond the proposed scope of Hepp in \cite{He}). On the other hand, we shall see that Sewell's approach is natural for that purpose.

The problem of irreversibility has been previously discussed in the literature by the introduction of various types of ''coarse graining'' : for beautiful discussions, see the articles by Lebowitz \cite{Leb} and Griffiths \cite{Gr}. There are strong arguments (see \cite{Gr} and section 5.3) that the ''arrow of time'' is provided by the direction of increase of the the quantum version of Boltzmann's entropy, the von Neumann macroscopic entropy (see \cite{Leb}, sec.8 and the introduction). Of course, if one wishes to relate the quantum Boltzmann approach to irreversibility to the theory of measurement, it is essential to include the measuring apparatus as part of the closed quantum mechanical system (see also the remarks in \cite{Gr}, pg.154).

The following version of the second law seems to encompass both observational (i.e. measurement) and thermal phenomena (theorem 1). Although several mathematical tools have already been developed by Lindblad in a series of beautiful papers (\cite{Lind1}, \cite{Lind2}, \cite{Lind3}), which had precisely measurement theory in mind, we use them in a different direction. In section 5a we illustrate theorem 1 by the Coleman-Hepp model.

In a letter to Niels Bohr (28-1-1947), Wolfgang Pauli remarks: ''The discussions I had here with Stern (he left Z\"{u}rich a few days ago) concerned the quantitative side of the connection of the concepts of entropy and observation, a connection which, as we all agree, is of a very fundamental character''. In this section, we revisit this topic, with a view to try to illuminate the aforementioned connection.

\subsection{Definition of the quantum Boltzmann entropy}

We adopt the setting of section 3.5, and define the projectors occurring in (3.19):
$$
P_{\alpha} = \mathbf{1} \otimes \Pi_{\alpha} \mbox{ for } \alpha = 1, \cdots n
\eqno{(4.1)}
$$
By (3.24), when $A \in {\cal A}$ is measured on the system $S$, initially in a state of the composite system described by a density matrix $\rho$, the value $\phi(\alpha) \in [1,n]$ is obtained with probability $w_{\alpha} = Tr(\rho P_{\alpha})$, after which the state of the composite system is described by the density matrix
$$
\rho^{'}_{\alpha} \equiv \frac{P_{\alpha} \rho P_{\alpha}}{w_{\alpha}}
\eqno{(4.2.1)}
$$
Averaging over all possible outcomes of the experiment yields the state
$$
\rho^{'} = \sum w_{\alpha} \rho^{'}_{\alpha} = \sum_{\alpha} P_{\alpha} \rho P_{\alpha}
\eqno{(4.2.2)}
$$
i.e., a mixture of states in each of which $A$ has a definite value. We define the \textbf{quantum Boltzmann entropy}
$S_{QB}(\rho)$ of a state $\rho$ by
$$
S_{QB}(\rho) \equiv -k Tr (\rho^{'} \log \rho^{'})
\eqno{(4.3)}
$$
where $\rho^{'}$ is given by (4.2.2), provided $\rho^{'} \log \rho^{'}$ is of trace class.

Definition 4.3 is adopted by several authors in the field, see, e.g., Zurek in \cite{Zu1}. The transformation
$$
\rho \to \rho^{'} = \sum_{\alpha} P_{\alpha} \rho P_{\alpha}
$$
may be viewed as a loss of information contained in the non-diagonal terms
$$
P_{\alpha} \rho P_{\alpha^{'}} \mbox{ with } \alpha \ne \alpha^{'}
$$
in
$$
\rho = \sum_{\alpha,\alpha^{'}} P_{\alpha} \rho P_{\alpha^{'}}
$$

\subsection{Histories and decoherence}

A generalization of (4.2.2), viz. when a sequence of measurements is carried out, and letting a time evolution intervene between measurements, leads to the assignment to a sequence of ''events''
$$
P_{\alpha_{1}}(t_{1}) P_{\alpha_{2}}(t_{2}) \cdots P_{\alpha_{n}}(t_{n})
$$
( a ''history'', briefly written $\underline{\alpha}$ for the index set of the corresponding vector) a probability distribution $W$,
$$
W(\underline{\alpha}) = TrP_{\alpha_{n}}(t_{n})\cdots P_{\alpha_{1}}(t_{1})\rho P_{\alpha_{1}}(t_{1})\cdots P_{\alpha_{n}}(t_{n})
\eqno{(4.4.1)}
$$
where $\rho = \rho(0)$, over the set of histories, where the $P$ satisfy the relations (3.19.1),(3.19.2) (written for $\Pi_{\alpha}\leftrightarrow P_{\alpha}$). This framework was proposed independently in \cite{Gr1}, \cite{GH}, \cite{O}. Let
\begin{eqnarray*}
D(\underline{\alpha^{'}},\underline{\alpha})=\\
= Tr P_{\alpha^{'}_{1}}(t_{1}) \cdots P_{\alpha^{'}_{n}}(t_{n}) \rho P_{\alpha_{n}}(t_{n}) \cdots P_{\alpha_{1}}(t_{1})
\end{eqnarray*}
$$\eqno{(4.4.2)}$$

We have (see, e.g., \cite{Th4}):

\textbf{Definition 4} A history is said to \textbf{decohere} iff
\begin{eqnarray*}
D(\underline{\alpha^{'}},\underline{\alpha})=\\
= \delta_{\underline{\alpha^{'}},\underline{\alpha}} \rho _{\underline{\alpha}}
\end{eqnarray*}
$$\eqno{(4.5)}$$

A state is called decoherent with respect to the set $P_{\alpha }$ iff

$$
P_{\alpha^{'}} \rho(0) P_{\alpha} = 0 \mbox{ for all } \alpha \ne \alpha^{'}
\eqno{(4.6)}
$$
This implies $$Tr P_{\alpha '}\rho P_{\alpha }A=0 \quad \forall \alpha \neq \alpha '$$ and this is equivalent to $$[P_{\alpha },\rho ]=0 \quad \forall \alpha $$

In contrast to infinite systems (see definition 1) where there is no need to refer to a choice of projections,  decoherent mixed states (over the macroscopic observables) can be described by relations between the density matrix a
 $\rho_{m}$ and the projectors. They are of the form
$$
\rho_{m} = |\Psi)(\Psi|
\eqno{(4.7)}
$$
with
\begin{eqnarray*}
\Psi \equiv \sum \lambda_{\alpha} P_{\alpha} \Phi_{\alpha}\\
\mbox{ with } \sum_{\alpha}|\lambda_{\alpha}|^{2} = 1 \mbox{ and } \\
\Phi_{\alpha} \in {\cal H} \mbox{ for } \alpha \in [1,n]
\end{eqnarray*}
$$\eqno{(4.8)}$$
(see (3.13.3) for an example) - are such that
$$
\sum_{\alpha<\alpha^{'}} (P_{\alpha^{'}} \rho_{m} P_{\alpha} + P_{\alpha} \rho_{m} P_{\alpha^{'}}) \ne 0
\eqno{(4.9)}
$$
(see section 5.1 for examples).

The algebra at infinity ${\cal L}_{\Pi}$ of section (2), which defines the ''pointer positions'' in Hepp's formulation \cite{He}, may also be viewed as a main feature - besides the inevitable indeterminacies in the values of macroscopic quantities - which lies behind the definition (4.3) of the quantum Boltzmann entropy. This is, of course, not new, and basic to von Neumann's approach \cite{vN}.  The elements of ${\cal L}_{\Pi}$ have to be considered as  the only ones accessible to experiment, and, indeed, all information on microscopic quantities, such as the spin of one electron in the Stern-Gerlach experiment, are, in principle, derivable from them, provided we can manipulate the time evolution in the system in an appropriate way. Therefore the reduced description of (4.3) is also natural from a purely conceptual point of view.

As Lindblad remarks (\cite{Lind1}, pg. 314)), (4.3) may be viewed as an averaging over the relative phases between the subspaces $P_{\alpha}({\cal H}\otimes {\cal K})$.

\subsection{Relative entropy and its properties}

We shall need the concept of \textbf{relative entropy} $S(\rho_{1}|\rho_{2})$ (called by \cite{Lind1} conditional entropy, but actually corresponding to the finite-dimensional version of the concept introduced subsequently by Araki \cite{Araki} in the general context of states on von Neumann algebras) between two states (positive operators of trace one), which is defined by
$$
S(\rho_{1}|\rho_{2}) = k Tr(\rho_{1} \log \rho_{1} - \rho_{1} \log \rho_{2})
\eqno{(4.10)}
$$

This expression is well studied  \cite{OP}, \cite{Wehrl1}, \cite{K}. We summarize the results we need and refer to \cite{OP} for the proofs.

\textbf{Lemma 1}
\begin{eqnarray*}
S(\rho_{1}|\rho_{2}) \ge 0
\end{eqnarray*}
$$\eqno{(4.11.1)}$$
\begin{eqnarray*}
S(\rho_{1}|\rho_{2}) = 0 \mbox{ iff } \rho_{1} = \rho_{2}
\end{eqnarray*}
$$\eqno{(4.11.2)}$$
\begin{eqnarray*}
\mbox{ If } \lambda \rho_{1} \le \rho_{2} \mbox{ for some } \lambda \in (0,1) \mbox {then}\\
S(\rho_{1}|\rho_{2}) \le -k \log \lambda
\end{eqnarray*}
$$\eqno{(4.11.3)}$$
\begin{eqnarray*}
S(\lambda \rho _1+(1-\lambda )\rho _2|\lambda \sigma _1+(1-\lambda )\sigma _2)\leq \lambda S(\rho _1|\sigma _1)+(1-\lambda )S(\rho _2|\sigma _2)
\end{eqnarray*}
$$\eqno(4.11.4)$$
\begin{eqnarray*}
S(\rho_{1}. \gamma|\rho_{2}. \gamma ) \le S(\rho_{1}|\rho_{2})
\end{eqnarray*}
$$\eqno(4.11.5)$$
where $\gamma $ is a completely positive map, e.g. an imbedding. The last two inequalities are known aa joint concavity and monotonicity of the relative entropy, both follow from Lieb's concavity theorem \cite{Liebb}. We shall denote von Neumann's entropy $S(\rho)= -k Tr(\rho \log \rho)$ simply by $S$ and call it ''the entropy''.

\subsection{A quantum version of the second law}

\textbf{Theorem 1}\, \textbf{A quantum version of the second law}

 Let the (initial) density matrix be assumed to be decoherent at zero time (4.6) with respect to $P_{\alpha }$ and to have finite entropy, i.e.,
$$\rho (0)=\sum _{\alpha }P_{\alpha }\rho (0)P_\alpha $$
$$
S_{QB}(\rho (0))=S(\rho(0))=- k Tr\rho (0) \log \rho (0) < \infty
\eqno{(4.12)}
$$
and assume it is not an equilibrium state of the system. Let $\rho(t_{0})$, $t_{0} > 0$, be any subsequent state of the system, possibly an equilibrium state. Then, for an automorphic (unitary) time-evolution of the system between $0 \le t \le t_{0}$,
$$
S_{QB}(0) \le S_{QB}(t_{0})
\eqno{(4.13)}
$$
with
\begin{eqnarray*}
S_{QB}(0) = S_{QB}(t_{0}) \mbox{ iff } \\
\sum_{\alpha < \alpha^{'}} P_{\alpha} \rho(t_{0}) P_{\alpha^{'}} + P_{\alpha^{'}} \rho(t_{0}) P_{\alpha} = 0
\end{eqnarray*}$$\eqno{(4.14)}$$

\textbf{Proof}\,
With $\rho'(t_0)=\sum _{\alpha }P_{\alpha }\rho (t_0)P_\alpha =\rho (t_0).\gamma $ where we indicate that $\rho '$ is obtained from $\rho $ by a completely positive map $\gamma$, we estimate
$$S(\rho '(t_0)|\rho '(0))=-S(\rho '(t_0))-\sum _{\alpha } k Tr(\rho (t_0)P_\alpha \log \rho (0) P_\alpha )= $$
$$-S_{QB}(t_0)- k Tr(\rho (t_0)\log \rho (0))=$$
$$\leq S(\rho (t_0)|\rho (0))=-S(\rho (0))-Tr (\rho (t_0)\log \rho (0))
\eqno{(4.15)}$$
In the first equality we used the cyclic property of the trace together with the definition of $\rho '$, in the second equality we used the decoherence of $\rho (0)$, the next inequality is a consequence of (4.11.5). In the last equality we use that the evolution is unitary and therefore preserves the entropy.
Together  this implies $$S_{QB}(t)\geq S_{QB}(0).$$. The equality condition in (4.14) follows from (4.11.2). q.e.d.

\subsection{Miscellaneous remarks: monotonicity in time, thermal and observational systems, Griffiths' ideas and the ''arrow of time''}

\textbf{Remark 1}\, The entropy growth in theorem 1 is not necessarily monotonic in the time variable. For this reason, we refer to \textbf{fixed} initial and final states in that theorem. For thermal systems, e.g., the adiabatic expansion of a gas discussed in the introduction, a natural choice of the final state is the equilibrium state of the gas. In the case of observational systems, it may be an incoherent (macroscopically) mixed state  such as the forthcoming (section 5a) density matrix of the Coleman-Hepp model for $t \ge t_{0}$, where $t_{0}$ is the ''decoherence time''.

\textbf{Remark 2}\, This remark contains some (at this point) entirely speculative observations related to theorem 1 and remark 1. Following an intuition due to Griffiths \cite{Gr} for thermal systems (eg. the irreversible expansion of a gas discussed in the introduction), the entropy growth is expected to be related to the property that, at time $t>0$, $P_{\alpha^{'}} \rho(t) P_{\alpha} \ne 0$ for some $P_{\alpha^{'}} > P_{\alpha}$; and, successively, $P_{\alpha^{''}} \rho(t^{'}) P_{\alpha^{'}} \ne 0$ with $P_{\alpha^{''}} > P_{\alpha^{'}}$ and $t^{'} > t$, that is, the macrostate will gradually be associated, with high probability, to subspaces of increasing dimension (in the simplified case that the latter are finite, also their number has to decrease), whereas $S_{QB}$ is still calculated with respect to the initial $P_{\alpha }.$ In this process, superpositions between states in different ${\cal H}_{\alpha}$ must be occurring all the time, leading to a final equilibrium entropy in agreement with theorem 1.

The above conjectures by Griffiths were formulated in an informal manner, with no claim to precision whatever. In the following we try to convey the logic of his formulation in a slightly more concrete manner, but much work will be necessary in order to find precise conditions under which this logic is correct, together with models which might serve as testing grounds.

The basic idea stems from an analogy with classical statistical mechanics, more specifically the fact that, as mentioned in (1.2), the totality of phase space points which at time $t_{0}$ were in a certain macrostate $M_{0}$ forms a region with fixed phase-space volume $|\Gamma_{M_{0}}|$ by Liouville's theorem, but whose shape will, for mixing systems, ''spread'' over regions $\Gamma_{M}$ corresponding to macrostates $M$ with increasingly large phase space volumes. Therefore, the probability that the macrostate is found in successively larger phase space volumes grows with time, and, with it, the classical Boltzmann entropy. 

Following \cite{Gr}, consider now a quantum mechanical system which at time $t_{0}$ is away from equilibrium, e.g., two interacting metal blocks at different temperatures which can exchange energy through a thin wire. Let a set of projections $\Pi_{\alpha}$ satisfying (3.19.1),(3.19.2) be defined, which represent a ''coarse graining'' in the sense that each projector $\Pi_{\alpha}$ specifies, within a reasonable tolerance, the energy of each of the metal blocks. Choose then an initial normalized state $|\Psi_{0})$ ''at random'' in the appropriate subspace of the Hilbert space corresponding to the specified initial energies. The corresponding $ D=|\Psi_{0})(\Psi_{0}|$ serves as initial condition upon which the histories with events at later times drawn from the set $\{\Pi_{\alpha}\}$ are conditioned. Due to the unitarity of the dynamics (the analogue of Liouville's theorem in the classical argument),
$$
Tr D(t) = ||\Psi_{0}(t)||^{2} = 1 = \sum_{j\in S_{t}} |c_{j}(t)|^{2}
\eqno{(4.16)}
$$
where $c_{j}(t)$ are coefficients associated to the expansion of $\Psi_{0}(t)$ in a fixed basis. A ''quantum ergodicity'' or ''quantum weak mixing'' (see later) implies that the cardinality of the set $S_{t}$ (supposed finite as in \cite{Gr}) increases with $t$, i.e, $\Psi_{0}(t)$ ''spreads'' (the analogue of the ''spreading '' of the initial volume of phase space in the classical case, keeping the phase space volume fixed). Therefore, the chances are that most of the probability in (4.16) (in the sense of (4.4.1)) will be associated to sequences $\Pi_{\alpha}$ with increasing dimensionality, implying the increase of the quantum Boltzmann entropy with time.

It should be remarked that even in the simplest cases, e.g., Laplacians or Schr\"{o}dinger operators on compact (or finite volume) Riemannian manifolds, with or without boundary, i.e., with ''chaotic'' geodesic flow, it does not seem easy to prove a conjecture of the type of the one formulated in the previous paragraph. If $\Psi_{0}$ is an eigenfunction of the Laplacian on this manifold, and the $c_{j}(t)$ in (4.16) are the coefficients of $\Psi_{0}(t)$ in the basis of the eigenfunctions of the same operator, the ''spreading '' has been proved in a certain precise, but semiclassical, sense - see the results on ''ergodic eigenfunctions'' and ''quantum weak mixing'' in \cite{Zel} and references given there. 

Concerning the possible non-monotonicity of the quantum Boltzmann entropy in time mentioned in remark 1, Griffiths remarks (\cite{Gr}, pg. 152) that it is crucial in the probability Ansatz he uses (for details, see \cite{Gr}) that it refers to the beginning $t_{0}$ of the period of time in which the system's entropy increases. Choosing a microstate at random within the phase space cell corresponding to the microstate reached by the original system at a certain intermediate time $t_{1} > t_{0}$ and integrating the classical equations of motion backwards in time, the result would - with overwhelming probability - \textbf{not} agree with the typical behavior of the entropy with the analogous specification at time $t_{0}$. ''To put it another way, statistical mechanics provides an explanation of an irreversible phenomenon by a probabilistic hypothesis which itself singles out a (thermodynamic) direction of time''. The nonmonotonic feature may be related to van Kampen's suggestion \cite{NGvK} that an additional assumption, e. g. of repeated random phases analogous to Boltzmann's molecular chaos, is indispensable for a theory of irreversibility. Indeed, theorem 1 suggests the possibility of a behavior of ''average growth'', similar to the analogous conjectured behavior for the Boltzmann H function (see, e.g., Fig. 1.4 in the elementary but very clear introduction in chapter 1 of \cite{CJT}).

Finally, what determines the ''arrow of time''? Griffiths \cite{Gr} remarks that, since two bodies at unequal temperatures but in thermal contact are known to exchange energy in such a way that the temperatures approach each other even in the presence of a magnetic field, which breaks time-reversal invariance, the latter is certainly not the key for understanding macroscopic irreversibility (see also the pioneer paper of Aharonov, Bergmann and Lebowitz \cite{ABL}, reprinted in \cite{WhZu}, as well as the discussion in \cite{Maes1} in the framework of classical statistical mechanics). Like Griffiths \cite{Gr}, we believe that it is the growth of the Boltzmann entropy that determines the arrow of time: in quantum theory, it points from an event away from equilibrium towards a situation of higher quantum Boltzmann entropy (compare also the discussion in (\cite{Haag}, second paragraph of page 312). For the role of the arrow of time in relativistic quantum field theory, see \cite{Bu}.

Some model results on the monotonic- versus - non-monotonic behavior of the entropy as a function of time are given in section 5.8.

\section{ Models of decoherence}

In spite of the existence of an enormous literature on decoherence and the transition from quantum to classical (see the books \cite{Giu} \cite{Pe}, the review articles \cite{Zu1} \cite{Zu2}, \cite{Wightman2}, there have been few papers dedicated to an exact purely quantum approach to the problem: notable are the aforementioned \cite{NTh1}, \cite{Th4}, see also previous work by Narnhofer \cite{Na} and Narnhofer and Robinson \cite{NRo}, the recent three dimensional model of decoherence induced by scattering \cite{CCF2}, and the exact quantum Brownian model of Unruh and Zurek \cite{UZ} and Joos and Zeh \cite{JZ}. The reason is, of course, the difficulty in obtaining some detailed information on the quantum evolution.

In sections 5.1 to 5.7 we illustrate theorem 1 by a specific model, a finite version treated by Sewell (\cite{Sew1}, \cite{Sew2}) of the Coleman-Hepp model introduced in section 3.2, which we extend to infinite space and times. In section 5.8 we look into models of the behavior of the quantum Boltzmann entropy for intermediate times,, introducing interactions between the chain spins. In sections 5.9 et seq. we consider a class of models of quantum chaotic systems (quantum Anosov systems) as models of decoherence.

\subsection{Application of theorem 1 to the Coleman-Hepp model}

We consider the finite Coleman-Hepp model , whose infinite version was introduced in section 3.2, following closely the approach of Sewell and the corresponding notation described in 3.5. According to (3.20) and (3.2), we take the Hilbert space of $S$ to be
$$
{\cal H} = L^{2}[-r,r] \otimes \mathbf{C}_{0}^{2}
\eqno{(5.1)}
$$
and
$$
H = \mathbf{1}_{{\cal H}} \otimes p
\eqno{(5.2)}
$$
where $p$ is the self-adjoint momentum operator on $L^{2}[-r,r]$, with some boundary condition
$\Phi(x+2r) = \exp(i\theta) \Phi(x)$, $\theta \in \mathbf{R}$, and
$$
V = (P_{-})_{0} \otimes \sum_{n=1}^{2L+1} v(x-n) \otimes \sigma_{n}^{1}
\eqno{(5.3)}
$$
Now the chain has a finite number $N=2L+1$ of spins and $n=2$ ; the index $1,2$ will be replaced by $\pm$ corresponding to the eigenvalues of $\sigma_{0}^{3}$. We define as in \cite{Sew1},\cite{Sew2} the macroscopic phase cells ${\cal K}_{\alpha}$ of section 3.5, now denoted ${\cal K}_{\pm}$, as the subspaces of ${\cal K} = \otimes \mathbf{C}^{2}_{n}$ spanned by the eigenvectors of the ''polarization operator''
$$
\Sigma_{L}^{3} \equiv \sum_{n=1}^{2L+1} \sigma_{n}^{3}
\eqno{(5.4)}
$$
with positive (resp. negative) eigenvalues; let $\Pi_{\pm}$ denote the associate projection operators. Of course, the restriction to these phase cells corresponds to an enormous reduction of information.
we can consider the time evolution for steps. It becomes
$$
\tau ^n P(\Phi )P(\Psi _0)\otimes _{l=1}^{2L+1}P(c_{l+}|+)+c_{l-}|-))=
$$
$$
=c_{0+}P(\Phi(x-n))\otimes |+)\otimes _{l=1}^{2L+1} P(c_{l+}|+)+c_{l-}|-))+
$$
$$
+c_{0-}P(\Phi(x-n))\otimes |-)\otimes _{l=1}^n P(c_{l+}|-)+c_{l-}|+))\otimes _{l=n+1}^{2L+1}P(c_{l,+}|+)+c_{l-}|-))$$
$$
\tau _x^k P(\Phi )=P(\Phi (x-k))
$$
$$
\tau ^{2L+1+k}=\tau _x^k \times \tau ^{2L+1}
\eqno{(5.5)}
$$

The model thus defined may (and will) be extended to infinite space (and time) by simply assuming the values taken at the boundary of the chain to be stationary (corresponding to $t_{\infty} = t_{0} = 2L+1-w+r$, see (3.15)): we interpret this as a statement that at this point and time the interaction with the chain ceases. Since the free Hamiltonian is a shift, no problem with unitarity arises. As remarked after (3.15), this yields a physically reasonable value of $t_{0}$ for a suitable choice of $L$ , and from this point of view is thus preferable to Hepp's choice of limits. In the rest of the paper, when we speak of the Coleman-Hepp model, we shall be referring to this version.

The correspondence between the phase cells and the eigenstates of $S$ in (3.24 2) is given by the mapping $r \to a(r)$ with
$a(\pm) = \pm$; i.e., the phase cells ${\cal K}_{\pm}$ are the indicators for the vector states $u_{\pm} = |\pm)_{0}$, with
$\sigma_{0}^{3} |\pm)_{0} = \pm |\pm)_{0}$. The $M_{\alpha}$ in (3.19.3) are the eigenvalues of $\Sigma_{L}^{3}$ in
${\cal K}_{\pm}$.

We consider two situations, characterized by two distinct types of initial state (3.21.1): one which is pure (over the microscopic states), originally explicitly considered in \cite{He};  another which is \textbf{not} pure over the microscopic states, but is decoherent at zero time, complying with the hypothesis of theorem 1, which was treated in \cite{Sew1}, \cite{Sew2}. We recall definitions (4.1) and (4.3), where, now, $\alpha = \pm$. We shall take
$$
\rho(0) = P(\Phi) \otimes P(\Psi_{0}) \otimes \Omega
\eqno{(5.6)}
$$
where
$$
P(\Phi) = |\Phi)(\Phi|
\eqno{(5.7.1)}
$$
$$
P(\Psi_{0}) = |\Psi_{0})(\Psi_{0}|
\eqno{(5.7.2)}
$$
$$
\Psi_{0} = c_{+} |+)_{0} + c_{-} |-)_{0}
\eqno{(5.7.3)}
$$
$\Omega$ is one of the following density matrices:
$$
\Omega_{\pm,L}(\beta = \infty) = 2^{-(2L+1)} \otimes_{=1}^{2L+1}(\mathbf{1}_{n} \pm \sigma_{n}^{3})
\eqno{(5.8.1)}
$$
\begin{eqnarray*}
\Omega_{+,L}(\beta < \infty) = 2^{-(2L+1)} \otimes (\mathbf{1}_{n} + m(\beta) \sigma_{n}^{3})\\
= \frac{\exp(\beta B \Sigma_{L}^{3})}{Z}
\end{eqnarray*}$$\eqno{(5.8.2)}$$
\begin{eqnarray*}
\Omega_{-,L}(\beta < \infty) = 2^{-(2L+1)} \otimes_{n=1}^{2L+1} (\mathbf{1}_{n} - m(\beta) \sigma_{n}^{3})\\
= \frac{\exp(-\beta B \Sigma_{L}^{3})}{Z}
\end{eqnarray*}$$\eqno{(5.8.3)}$$
where
$$
m(\beta) = \tanh(\beta B)
\eqno{(5.8.4)}
$$
Above, $\Sigma_{L}^{3}$ is given by (5.4),
$$
Z = Tr_{{\cal K}} \exp(\beta B \Sigma_{L}^{3}) = Tr_{\cal K} \exp(- \beta B \Sigma_{L}^{3})
\eqno{(5.8.5)}
$$
is the partition function, and
$$
\pm m(\beta) = \frac{Tr_{{\cal K}}(\Omega_{\pm,L}(\beta < \infty) \Sigma_{L}^{3})}{2L+1}
\eqno{(5.8.6)}
$$
is the macroscopic polarization.

As noted by Sewell (\cite{Sew1},\cite{Sew2}), since the initial state $\Omega_{\pm,L}(\beta = \infty)$ is transformed into
$\Omega_{\pm,L}(\beta < \infty)$ by small perturbations of the global polarization $m$, it is natural to consider both, and, thus, only the ''apparatus'' corresponding to finite temperature is stable under such small perturbations - a natural requirement on any measuring apparatus.

In \cite{Sew1} , \cite{Sew2} values different from the value one-half for the quantity (3.4.2) were considered, but, except for some remarks in connection with typicality, we shall take $J = 1/2$ as in (3.4.2). This justifies to consider the discrete time evolution (5.5).

\subsection{The microscopically pure case $\beta = \infty$}

In this case $\rho(0)$ is pure (over the microscopic states)  given by (5.6) and (5.8.1). Hypothesis (4.12) of theorem 1 is thus true a fortiori. By (4.1) and (4.3), we have the general form:
$$
S_{QB}(\rho) = -k Tr(P_{+} \rho P_{+} \log P_{+} \rho P_{+}) - k Tr(P_{-} \rho P_{-} \log P_{-} \rho P_{-})
\eqno{(5.9)}
$$
where the trace is over ${\cal H} \otimes {\cal K}$. We shall denote any $t$ satisfying (3.10.4) for $n = 2L+1$ by
$t = \infty$. By (5.7),
$$
\rho(0) = |\Phi)(\Phi| \otimes |\Psi_{0})(\Psi_{0}| \otimes_{i=1}^{2L+1} |+)_{i}(+|_{i}
\eqno{(5.10)}
$$
with a slight abuse of Dirac notation, in which we shall be incurring in what follows. By (5.10), and defining
$$
\rho_{\pm} \equiv P_{\pm} \rho P_{\pm}
\eqno{(5.11)}
$$
we have
$$
\rho_{+}(0) = \rho(0)
\eqno{(5.12.1)}
$$
and $$
\rho_{-}(0) = 0
\eqno{(5.12.2)}
$$
By (5.12),
$$
S_{QB}^{\beta = \infty}(\rho(0)) = 0
\eqno{(5.13)}
$$
Consider,now, $t=\infty$ (i.e. $2L+1<t<<r$). Define
\begin{eqnarray*}
\tilde{\Psi_{t}} \equiv |\Phi_{t})\otimes\
(c_{+} |+)_{0} \otimes_{i=1}^{(2L+1)} |+)_{i} + c_{-} |-)_{0} \otimes_{i=1}^{(2L+1)}|-)_{i})
\end{eqnarray*}$$\eqno{(5.14.1)}$$
where
$$
\Phi_{t}(x) = \Phi(x-t)
\eqno{(5.14.2)}
$$
from which
$$
P_{+} \tilde{\Psi_{t}} = |\Phi_{t}) \otimes (c_{+} |+)_{0} \otimes_{i=1}^{(2L+1)} |+)_{i})
\eqno{(5.15.1)}
$$
$$
P_{-} \tilde{\Psi_{t}} = |\Phi_{t}) \otimes (c_{-} |-)_{0} \otimes_{i=1}^{(2L+1)} |-)_{i})
\eqno{(5.15.2)}
$$
and
\begin{eqnarray*}
\rho^{\beta = \infty}(\infty) = |\Phi_{t})(\Phi_{t}| \otimes \
(|\tilde{\Psi}_{t})(\tilde{\Psi}_{t}|
\end{eqnarray*}
$$\eqno{(5.15.3)}$$
where the pointer projectors correspond to
$$
\tilde{P}_{\pm} = \mathbf{1}_{\mathbf{C}_{0}^{2}} \otimes \Pi_{\pm}
\eqno{(5.15.4)}
$$
By (5.15.1,2),
\begin{eqnarray*}
P_{+} \rho^{\beta = \infty}(\infty) P_{+} + P_{-} \rho^{\beta = \infty}(\infty) P_{-}\\
= |\Phi_{t})(\Phi_{t}| \otimes (|c_{+}|^{2} |+)_{0}(+|_{0} \otimes_{i=1}^{(2L+1)}|+)_{i}(+|_{i}+\\
+ |c_{-}|^{2} |-)_{0}(-|_{0} \otimes_{i=1}^{(2L+1)}|-)_{i}(-|_{i})
\end{eqnarray*}$$\eqno{(5.16)}$$
Putting (5.16) into (5.9),
$$
S_{QB}^{\beta = \infty} (\rho(\infty)) = -k |c_{+}|^{2} \log |c_{+}|^{2} - k |c_{-}|^{2} \log |c_{-}|^{2}
\eqno{(5.17)}
$$
Notice the cross terms in (5.15.3)
\begin{eqnarray*}
Q_{1} \equiv c_{+}\bar{c_{-}} |+)_{0}(-|_{0} \otimes_{i=1}^{(2L+1)} |+)_{i}(-|_{i} +\\
+ c_{-}\bar{c_{+}} |-)_{0}(+|_{0} \otimes_{i=1}^{(2L+1)}|-)_{i}(+|_{i}
\end{eqnarray*}$$\eqno{(5.18)}$$
which are such that $P_{+} Q_{1} P_{-} \ne 0$.
The effect is stable under perturbations of the initial state:

\subsection{The mixed case $\beta < \infty$}

In the case $\beta < \infty$ and taking $\omega = \Omega_{+,L}(\beta < \infty)$ in (5.6), we see that $\rho(0)$ is  mixed state, but (5.8.2) shows explicitly that $\rho$ is decoherent at zero time, because $[P_{\pm},\rho ]=0.$ The density matrix at $t=0$ is given by
$$
\rho_{L}^{\beta < \infty}(t=0) = P(\Phi) \otimes P(\Psi_{0}) \otimes \Omega_{+,L}
\eqno{(5.19)}
$$
Let $E_{i}$ denote the (degenerate) eigenvalues of $\Sigma_{L}^{3}$ on the basis $\{|i)\}_{i=1}^{2^{(2L+1)}}$ of eigenstates of
$\otimes_{i=1}^{2L+1} \sigma_{i}^{3}$. These eigenstates can be characterized by partitions of the set $\{l=1,.... =2L+1 \}=I\bigcup J$
so that

$$
\Psi_{\pm,I} =\otimes _{l\in I}|\pm)_l\otimes _{k\in J}|\mp)_k
\eqno{(5.20)}
$$

These vectors according to (5.5) are mapped at $t_{\infty}$ into

$$|c_{0+}|+)_0+c_{0-}|-)_0 \otimes\Psi_{\pm,I}\rightarrow $$
$$c_{0+}|+)_0 \otimes \Psi_{\pm,I}+c_{0-}|-)_0 \otimes \Psi _{\mp,I}
\eqno(5.21)
$$
The projection operators commute with $P(\Psi_{\pm,I})$ so that the density operator remains decoherent, but now with different weights for the contribution, namely
 $$ \rho = P_+\rho P_+ +P_-\rho P_-=$$
 $$=\sum _I (c_+^2\alpha _I^2 +c_-^2 \alpha _{I^c}^2)|\Psi _I^+\langle \rangle \Psi _I^+ |+$$
 $$+\sum _I (c_-^2\alpha _I^2 +c_+^2 \alpha _{I^c}^2)|\Psi _I^-\langle \rangle \Psi _I^-|$$
 As for the ground state the coherence of the system that is measured turned into a decoherence of the apparatus so that
$$
S_{QB}(\rho (t_{\infty }))= S(c_+^2\rho _{\beta }+c_-^2 \rho _{-\beta })
\eqno{(5.22)}
$$
Notice that the initial coherence of the state of the measured system turned into decoherence of the apparatus, but with a weight that depends not only on the coherence but also on the initial state of the apparatus. This is a feature that we do not expect from the pointer position of a classical measurement.

\subsection{The ''bad'' microstates}

\textbf{Remark 3} In the classical theory discussed in the introduction there arise ''bad microstates'', e.g., those microstates consisting of gas molecules whose velocity vectors are directed away from the barrier lifted at time $t_{0}$. The analogous states in the Coleman-Hepp model are just those leading to nonzero ''nondiagonal'' terms
$\Pi_{-} \Omega_{+,L}$ and $\Pi_{+} \Omega_{-,L}$, which hamper the approach of the coherent mixture (5.21) to the decohered or incoherent mixture
$$
|c_{+}|^{2} P_{+}^{0} Tr( \Omega_{+,L} .) + |c_{-}|^{2} P_{-}^{0} Tr(\Omega_{-,L} .)
$$
The fact that their contribution is ''small'' is measured by
$$
m_{+} \equiv Tr (\Pi_{-} \Omega_{+,L}(\beta < \infty)) = O(\exp[-c(2L+1)])
\eqno{(5.23.1)}
$$
and
$$
m_{-} \equiv Tr (\Pi_{+} \Omega_{-,L}(\beta < \infty)) = O(\exp[-c(2L+1)])
\eqno{(5.23.2)}
$$
Above,
$$
c = \log \cosh(\beta B)
\eqno{(5.23.3)}
$$
(5.23) follow follow directly from the second formulae in (5.8.2,3).

\subsection{Reduction of the wave packet according to definitions 2 and 3}

From the point of view of the reduction of the wave-packet, we have:

\textbf{Proposition 1} Reduction of the wave packet in the sense of definition 2 occurs for both $\beta = \infty$ and
$\beta < \infty$. Reduction of the wave packet in the sense of definition 3 occurs for $\beta = \infty$ for any
$L \ge 0$ and, for $\beta < \infty$, but here only in the limit $L\rightarrow \infty $ for all $A$ or for all $L$ but $[A,P_+]=0.$

\textbf{Proof} We have
$$\langle \Psi_{+,I}|\Psi _{-,I}\rangle =0$$
For $\beta =\infty $ only $I=\{1... 2L+1\}$ contributes. For $\beta <\infty $ we have to take into account the contribution of
$$\langle \Psi_{+,I}|A|\Psi _{-,I}\rangle $$
This contribution vanishes for all $I$ that are larger than the localization of $A$, therefore for almost all $I$ that contribute to the expectation value with $\rho (t_{\infty })$ if $L$ tends to $\infty.$ More precisely
$$
\langle \Psi_{+,I}|A|\Psi _{-,I}\rangle <O(exp[-c(2L+1)])
\eqno{(5.23.4)}
$$
as explained in remark 3. q.e.d.

\textbf{Remark 4} The case $\beta = \infty$ in proposition 1 exemplifies the fact that definition 3 - in contrast to definition 2 - does not rely a priori on the apparatus ${\cal I}$ having a large number of degrees of freedom - see also the comments after definition 2. The case $L=0$, i.e., only one degree of freedom, is taken over in section 5c (see definition 5 there), in a model of decoherence, and seems to agree well with the ideas of Zurek and collaborators, expounded in the introduction and in that chapter, that only one degree of freedom in a chaotic environment may be more effective for decoherence that the quantum Brownian motion models with reservoirs with a large number of degrees of freedom.

On the other hand, the case $\beta = \infty$ treated by Hepp is very special, and Sewell showed that natural stability requirements imposed on the apparatus lead to consider the mixed case $\beta < \infty$, for which the number of degrees of freedom being large is \textbf{crucial}, being related to the exponentially small corrections mentioned in proposition 1.

It is to be noted that $\rho^{\beta<\infty}(t=\infty)$ contains  terms
$$
Q_{2} = c_{+}\bar{c_{-}}|+)_{0}(-|_{0} \otimes |\Psi_{+,I})(\Psi_{-,J}|
+ \bar{c_{+}}c_{-} |-)_{0}(+|_{0} \otimes |\Psi_{-,J})(\Psi_{+,I}|
\eqno{(5.24.1)}
$$
with
$$
P_+|\Psi _{+,I})=|\Psi _{+,I}), \quad P_-|\Psi_{-,J})=|\Psi_{-,J})
$$
As a consequence

\textbf{Lemma 2} For $L$ large but finite
$$
P_{+} \rho^{\beta<\infty}(t=\infty) P_{-} + P_{-} \rho^{\beta< \infty} P_{+} \ne 0
\eqno{(5.25.1 )}
$$
\textbf{Proof} Let
$\Theta_{1} \equiv |+)_{0} \otimes |\Psi_{+,L})$, $\Theta_{2} \equiv |-)_{0} \otimes |\Psi_{-,L})$, and denote by
$D_{\pm} \equiv |\Psi_{\pm,L})(\Psi_{\pm,L}|$ the diagonal terms in $\rho^{\beta < \infty}$. We have that
$\Theta_{1} \ne 0$ and $\Theta_{2} \ne 0$ (they are vectors of norm one each), and the following hold, for some $c > 0$:
$$
(\Theta_{1}, Q_{2} \Theta_{2}) = c_{+} \overline{c}_{-} + \mbox{ terms of order } O(\exp[-c(2L+1)])
\eqno{(5.25.2)}
$$
$$
|(\Theta_{1}, P_{-} \rho^{\beta< \infty} P_{+} \Theta_{2})| = O(\exp[-c(2L+1)])
\eqno{(5.25.3)}
$$
$$
(\Theta_{1}, D_{\pm} \Theta_{2}) = O(\exp[-c(2L+1)])
\eqno{(5.25.4)}
$$
This follows by the same procedure outlined in remark 3. (5.25.2-4) imply (5.25.1). q.e.d.

Thus $\rho^{\beta<\infty}(t=\infty)$ is (macroscopically) mixed . Theorem 1 is applicable to this more general case and yields the expected growth of the quantum Boltzmann entropy, here nontrivial because a direct treatment is hindered by the cross terms $Q_{2}$ (5.24), which do not disappear as in the case $\beta = \infty$ by forming
$P_{+} Q_{2} P_{+} + P_{-} Q_{2} P_{-}$. However, again as in remark 3,
$$
|Tr(P_{+} Q_{2} P_{+} + P_{-} Q_{2} P_{-})| = \mbox{ O } (\exp[-c(2L+1)])
\eqno{(5.25.5)}
$$
 We summarize:

\textbf{Proposition 2} For any initial state (5.7.3) such that $c_{+} c_{-} \ne 0$, we have: if  $\beta = \infty$,
\begin{eqnarray*}
\Delta S_{QB}^{\beta = \infty} \equiv S_{QB}^{\beta = \infty} (\rho(\infty)) -S_{QB}^{\beta = \infty} (\rho(0)) =\\
= -k (|c_{+}|^{2} \log |c_{+}|^{2} + |c_{-}|^{2} \log |c_{-}|^{2}) > 0
\end{eqnarray*}$$\eqno{(5.26.1)}$$. For $0 < \beta < \infty$,
$$
\Delta S_{QB}^{\beta < \infty} \equiv S_{QB}^{\beta < \infty} (\rho(\infty)) -S_{QB}^{\beta < \infty} (\rho(0)) > 0
\eqno{(5.26.2)}
$$

\textbf{Proof} (5.26.1) follows from (5.13) and (5.17), (5.26.2) from (4.14) of theorem 1 and Lemma 2. q.e.d.

\subsection{Miscellaneous remarks: the role of the quantum Boltzmann entropy in the vanishing of cross-terms, and the notion of temperature}

\textbf{Remark 5}\, It is interesting that, by definition (4.3) of the quantum Boltzmann entropy, the cross-terms $Q_{1}$ and $Q_{2}$ are either vanishing or small: therefore, the density matrix behaves, in a sense, as a coherent mixture (with (5.26.1) as expected result), and the reduction of information in the definition of the quantum Boltzmann entropy eliminates the cross-terms already for finite systems and times.

For $\beta < \infty$ the positive term on the r.h.s. of (5.26.2) may be extensive , but this extensive term tends to zero, as $\beta \to \infty$.

\textbf{Remark 6}\, It should be emphasized that $\beta < \infty$ and $\beta = \infty$ are suggestive of the type of state we considered, but has nothing to do with ''inverse temperature of the system'', a notion reserved for equilibrium states. If the infinite-volume version of the state described in theorem 1 is a stationary state (e.g. for $t = \infty$ in Hepp's formulation of the Coleman-Hepp model) with the property of passivity \cite{PuWo}, it follows \cite{PuWo} that it is either a ground state or a temperature (KMS) state, and that the temperature parameter defined in the latter case may indeed be identified with the absolute temperature of thermodynamics.

\subsection{Entropy reduction in events causing purification by state collapse and the role of quantum chaotic systems (quantum K systems) in the purification process}

\textbf{Remark 7}\, The model with($\beta < \infty$) provides an example of a density matrix at zero time which is a mixed state but decoheres. By proposition 1, its weak limit, as $L \to \infty$, (represented by state collapse defined by a family of projectors, i.e. the final state is macroscopically pure according to the given probability)
 is long enough, these events purify the mixed state on the classical quantities. In the case $n=2$ in (2.16.1), we have
$$
\omega =W_{\omega}(\underline{\alpha})\omega _{(\underline{\alpha})}
$$
The authors of \cite{NTh1} prove that for almost all histories (i.e. $\underline{\alpha }$) \textbf{either}
 $$
 \lim _{\underline{\alpha}}\omega_{\underline{\alpha}}=\omega _1
 \eqno{(5.27.1)}
 $$
 \textbf{or}

 $$
 \lim _{\underline{\alpha}}\omega_{\underline{\alpha}}=\omega _2
 \eqno{(5.27.2)}
 $$
 such that either $\omega _1$ or $\omega _2$ dominates, but no mixture of them.  One cannot dominate over all histories because
$$
\sum_{\underline{\alpha}} W_{\omega_{1}}(\underline{\alpha}) = \sum_{\underline{\alpha}} W_{\omega_{2}}(\underline{\alpha}) = 1
\eqno{(5.27.3)}
$$
For this reason, macroscopic purification refers to an \textbf{undetermined} individual outcome, and it happens under the assumption of collapse of the state at each individual measurement.

Quantum K systems are closest to classical chaotic systems in the sense that the time correlations factorize when the differences in the time arguments become very long. to the extent that most Hamiltonians are quantizations of classical chaotic Hamiltonians in a generic classical sense, the result of \cite{NTh1} is a partial explanation of the macroscopic purity found in Nature.

On the other hand, there are problems with ''what happens if things are not measured'': Griffiths \cite{Gr} remarks that a colleague once asked him, half-seriously, at a conference, ''Is Jupiter really there when no one has his telescope pointed in that direction?''. Concerning this last question, one may quote the following from Narnhofer and Thirring \cite{NTh1}, which relates to the observables assuming definite values in e.g. (2.12.1). They observe that, e.g. below the phase transition, domains of a magnet will be magnetized in a definite direction - for instance, the value of the mean magnetization (2.15) for some \textbf{fixed} $\vec{n}$ is obtained. ''Even if nobody looks at them, there will be enough 'events'-i.e., interactions with the environment - to purify the state over the classical part''.

It is important to stress that the relation between the above considerations and the main text concerns \textbf{only} the preliminaries, that is to say, as indications that the assumption in theorem 1 that $\rho$ decoheres at zero time is a natural one. In fact, the events referred to above act to decrease our uncertainty regarding the state, and may be expected, in
general, to lead to a \textbf{decrease} of the quantum Boltzmann entropy:

\textbf{Lemma 3 } The state collapse of a decoherent state produces ,in the average, a reduction of the quantum Boltzmann entropy.

\textbf{Proof} In a state collapse the state corresponding to the density matrix $\rho $ becomes one of the states corresponding to $\rho _i = (Tr\rho P_i)^{-1}P_i\rho P_i$ with probabilities $Tr \rho P_i$. Therefore in the average the entropy will be
$$ \sum _i (Tr\rho P_i)S(\rho _i ) =\sum _i (Tr\rho P_i)S((Tr\rho P_i)^{-1}\rho ^{1/2} P_i \rho ^{1/2}) < $$
$$S(\rho )
\eqno{(5.28)}
$$
Here we used the fact that the spectrum of $AA^{\dag}$ equals the spectrum of $A^{\dag}A$ and the strict concavity property of the entropy. Replacing $\rho$ above by $\sum P_{i} \rho P_{i}$ the same inequality for the quantum Boltzmann entropy follows. q.e.d.

\subsection{Monotonic versus non-monotonic behavior of the quantum Boltzmann entropy as a function of time: some models}

In the previous model we started with a product state on the microsystem $S$ and the apparatus $\cal I.$ By interaction the state turned immeadetly into an entangled state so that the coherence of the initial state on $S$ turned into coherence of the combined system. Reduction of the state to the apparatus gives decoherence with respect to some projections, that however will in general not be macroscopic. Continuing interaction between $S$ and ${\cal I }$ will make them macroscopic, but this interaction makes it necessary that the microsystem itself is of infinite (or at least large) size, (i.e. $r>>t$).

Another possibility is an interaction inside of the apparatus, such that the interaction between $S$ and ${\cal I } $
produces again a coherent state that evolves only in ${\cal I}$ to a decoherent state with respect to macroscopic observables. This model is inspired by an apparatus where a photon sets an electron free that itself will produce an avalanche of other photons.
We take for $S$ a spinsystem that interacts with a tensor product of four spin systems that in addition can move.
We start with the initial state
$$\Psi =(c_+|+)+c_-|-))\otimes |1+,2+,3+,4+)$$
The interaction between $S$ and ${\cal I}$ changes the state to
$$\Psi =(c_+|+)\otimes |1+,2+,3+,4+)+c_|-)\otimes |1-,2+,3+,4+)$$
Now the spins in ${\cal I}$ move and interact such that the states evolve
$$\Psi =(c_+|+)\otimes |1+,2+,3+,4+)+c_-|-)\otimes |1-,2+,3+,4+)$$
$$\Psi =(c_+|+)\otimes |1+,2+,3+,4+)+c_-|-)\otimes |1-,2-,3+,4+)$$
$$\Psi =(c_+|+)\otimes |1+,2+,3+,4+)+c_-|-)\otimes |2-,3-,4-,1+)$$
$$\Psi =(c_+|+)\otimes |1+,2+,3+,4+)+c_-|-)\otimes |3-,4+,1-,2-)$$
$$\Psi =(c_+|+)\otimes |1+,2+,3+,4+)+c_-|-)\otimes |4+,1+,2-,3-)$$
$$\Psi =(c_+|+)\otimes |1+,2+,3+,4+)+c_-|-)\otimes |1+,2+,3-,4+)$$
$$\Psi =(c_+|+)\otimes |1+,2+,3+,4+)+c_-|-)\otimes |2+,3+,4-,1+)$$
$$\Psi =(c_+|+)\otimes |1+,2+,3+,4+)+c_-|-)\otimes |3+,4+,1-,2-)
\eqno{(5.29.1)}
$$

We observe that the total spin fluctuates but never reaches the maximal value. If we choose as "macroscopic" projections that the total spin is either positive or negative, the quantum Boltzmann entropy fluctuates.In addition in the average the total spin vanishes in the terms corresponding to $c_-.$
 The dimension of the Hilbert space is now $d=16,$ whereas
$$\dim \bigcup _lV^l|\Psi _-\rangle =6,\quad \dim P(-1)\bigcup _lV^l|\Psi _-\rangle =2 $$ $$\dim P(0)\bigcup _lv^l|\Psi _-\rangle =2 \quad \dim P(1)\bigcup _lV^l|\Psi _-\rangle =2$$
As other example we took $n=6.$ Now the dimension of the Hilbert space is $d=64.$ Otherwise
$$\dim \bigcup _lV^l|\Psi _-\rangle =30,\quad \dim P(-4)\bigcup _lV^l|\Psi _-\rangle =3 $$ $$\dim P(-2)\bigcup _lv^l|\Psi _-\rangle =3 \quad \dim P(0)\bigcup _lV^l|\Psi _-\rangle =12$$ $$\dim P(2)\bigcup _lv^l|\Psi _-\rangle =6 \quad \dim P(4)\bigcup _lV^l|\Psi _-\rangle =16$$
For other choice (e.g the permutations (254613) or (234516)) we obtain

$$\dim \bigcup V^l|\Psi _-\rangle =26, 31; $$
with mean magnetization in the course of time
$$\eta _l(\langle \Psi _-|V^l \sigma _zV^{-l}|\Psi _-\rangle ) = -8/26, -3/31$$
so that more chaotic permutations reduce the magnetization in the time average close to 0 though of course the magnetization fluctuates.

The model can be generalized to a larger number $n$ of spins correspondingly that ${\cal I}$ is taken to be macroscopic and to a more random movement of the spins. More precisely we consider a perturbation of the points $\Pi (j)$ and interaction between ordered pairs
 $$U\Phi ^-=V_{\Pi(j)}\otimes_k^{n/2} U_{2k-1, 2k}\Phi ^-$$
 mit
 $$U_{1,2}|1+,2+)=|1+,2+)$$
$$U_{1,2}|1-,2+)=|1-,2-)$$
$$U_{1,2}|1+,2-)=|1+,2-)$$
$$U_{1,2}|1-,2-)=|1-,2+)
\eqno{(5.29.2)}
$$
Such an interaction will at the beginning decrease the total spin. If the perturbation is sufficiently random also the pairing should be sufficiently random and hit with approximately equal probability pairs in a way that the total spin should change of order less than the number of all spins. With increasing number of spins we may hope that the vector corresponding to $c_-$ will lead to an expectation value $0$ for the total spin,  the vector corresponding to $c_+$ remains unchanged and will give an expectation value $1$ for the mean spin, which makes it possible to choose the projections on a macroscopic level. Of course the time evolution of a finite system is periodic, therefore we can expect monotonic increase of the quantum Boltzmann entropy and macroscopic decoherence only for some time interval according to the considerations of Sewell.

As for the Coleman-Hepp model we can make the apparatus infinite, i.e. take $n=\infty $ and replace the perturbation of the points by a movement where every particle moves to the next place, where the spin is not turned around, in the ordering in which they were hit. This corresponds to a shift in the Hilbert space. Since the Hilbert space is infinite, we do not worry, whether it is unitary. Therefore the number of turned spins will increase exponentially in time and the state will converge in the weak limit but nevertheless fast to a decoherent state. Comparing the Coleman-Hepp model with the models in 5.8 we realize that in both models we obtain immediately entanglement of the system with the apparatus and this entanglement is kept in the course of time. In the Coleman-Hepp model the passage to macroscopic observation is due to a macroscopic duration of the interaction between the system and the apparatus whereas in the other models the interaction between system and apparatus is instantaneous and the passage to macroscopic values happens only in the apparatus. however now this asks differently frome the Coleman-Hepp model for a precise initial state, that is only metastable, therefore exceptional for the otherwise random evolution of the apparatus.

\subsection{Decoherence for quantum Anosov systems}

In remark 7, we have dwelt on the important role of quantum K systems in promoting macroscopic purification. In this section we revisit a different class of quantum chaotic models - quantum Anosov systems - for which some exact results have been obtained, in particular an estimate for the time scale of decoherence.

\subsection{ Quantum Anosov systems: definitions, motivation and connection to classical mechanics} 

In classical statistical mechanics, ergodicity and mixing play a crucial role in the theory of approach to equilibrium (\cite{Pen}). Classical Anosov systems display these properties due to the phenomenon of trajectory instability (see, again, \cite{Pen} for a brief review and references), or sensitive dependence on initial conditions. The intuition behind the latter property is that next to any point on any trajectory there is a point at distance $s$ such that their distance grows exponentially in the course of time. \textbf{Quantum Anosov systems} have been introduced by Emch, Narnhofer, Sewell and Thirring \cite{ENST}, see also \cite{MK}, \cite{ViM}, \cite{JSGW} and \cite{SJWe}.  If $\tau_{t}$ is the time translation automorphism (of the algebra ${\cal A}$ as in section 2), and $\sigma_{s}$ the space translation automorphism,
$$
\tau_{t} \circ \sigma_{s} = \sigma_{s\exp(-\lambda t)} \circ \tau_{t}
\eqno{(5.30)}
$$
From (5.30) one abstracts the multiplication laws of the Anosov semigroup, if $(t,s) \in \mathbf{Z} \times \mathbf{R}_{+}$, Anosov group if $(t,s) \in \mathbf{Z} \times \mathbf{R}$, or continuous Anosov group if
$(t,s) \in \mathbf{R} \times \mathbf{R}$ \cite{Th2}. In the case of abelian C* algebras, i.e., spaces of continuous functions, we have the classical Anosov groups, of which a prototype is represented by repulsive harmonic forces, which displays sensitive dependence on initial conditions because the particle runs with exponentially increasing velocity to infinity. The classical Hamiltonian $H$ and the classical dilation $K$ are given by
$$
H = \frac{\lambda (p^{2} - x^{2})}{2}
\eqno{(5.31.1)}
$$
$$
K = p - x
\eqno{(5.31.2)}
$$
with corresponding automorphisms
$$
\tau_{t}(x,p) = (\cosh t x + \sinh t p, \cosh t p + \sinh t x)
\eqno{(5.31.3)}
$$
$$
\sigma_{s}(x,p) = (x + s, p + s)
\eqno{(5.31.4)}
$$
which satisfy (5.30). The one-particle quantum mechanics version of (5.30) \cite{Th2} may be written
$$
\exp[iKs\exp(-\lambda t)] \exp(iHt) = \exp(iHt) \exp(iKs)
\eqno{(5.32)}
$$
where $H$ and $K$ are assumed to be essentially self-adjoint on $C_{0}^{\infty}(\mathbf{R})$, and therefore generate automorphisms of ${\cal B}(L^{2}(\mathbf{R})$ by
$$
\tau_{t}(A) = \exp(iHt) A \exp(-iHt)
\eqno{(5.33.1)}
$$
$$
\sigma_{s}(A) = \exp(iKs) A \exp(-iKs)
\eqno{(5.33.2)}
$$
which satisfy (5.32). For the quantum version of (5.31), (5.31.1) and (5.31.2) hold with $x$ the multiplication operator and
$p = -i\frac{\partial}{\partial x}$, the momentum operator on $L^{2}(\mathbf{R})$. The algebra ${\cal A}$ is the Weyl algebra $W$ generated by
$$
W \equiv W(\beta,\gamma) = \exp[i(\beta x+\gamma p)]
\eqno{(5.34.1)}
$$
A more interesting example is the parametric quantum oscillator, one of the simplest paradigms of the transition from regular to unstable behavior in classical mechanics (see \cite{JSGW} and references given there). The Hamiltonian is (again with mass $m = 1$):
$$
H(t) = p^{2}/2 + f(t) x^{2}/2
\eqno{(5.35.1)}
$$
where $f$ is a periodic function of period $T$:
$$
f(t+T) = f(t)
\eqno{(5.35.2)}
$$
In this case, we define
$$
K_{\underline{\alpha}} \equiv \alpha_{p} p + \alpha_{x} x \quad
\mbox{ with } \underline{\alpha} \equiv (\alpha_{p},\alpha_{x}) \in \mathbf{R}^{2}
\eqno{(5.36)}
$$
which satisfies
$$
[K_{\underline{\alpha}}, W(\beta,\gamma)] = (\alpha_{p} \beta - \alpha_{x} \gamma) W(\beta,\gamma)
\eqno{(5.37)}
$$
We assume that the dynamics defines an automorphism of $W$,
$$
U^{\dag}(t,t_{0}) A U(t,t_{0}) \in W \mbox{ for all } A \in W \mbox{ and } t,t_{0} \in \mathbf{R}
\eqno{(5.38)}
$$
where $U(t,t_{0})$ is the unitary family of propagators associated to (5.35.1). 

\subsection{The upper quantum Lyapunov exponent}

Under the above assumptions, one may define \cite{JSGW} the \textbf{upper quantum Lyapunov exponent} $\bar{\lambda}$ as
$$
\bar{\lambda} = \sup_{\alpha \in \mathbf{R}^{2}} \bar{\lambda_{\underline{\alpha}}}
\eqno{(5.39.1)}
$$
where
$$
\bar{\lambda_{\underline{\alpha}}}(U,L_{\underline{\alpha}},A,t_{0}) \quad
\equiv \limsup_{t \to \infty} \frac{\log||[L_{\underline{\alpha}}, A(t,t_{0})]||}{t}
\eqno{(5.39.2)}$$
where
$$
A(t,t_{0}) \equiv U^{\dag}(t,t_{0}) A  U(t,t_{0})
\eqno{(5.39.3)}
$$
and $A \in W$. The norm, as in section 2, is $||A|| = \sup_{\Psi \in {\cal H}}||A \Psi||/|| \Psi||$.

Example (5.31) may be considerably extended to include the case of an almost periodic quantum parametric oscillator \cite{SJWe}, i.e., for which $f$ in (5.35.1) is periodic (5.35.2), quasi-periodic or almost-periodic; accordingly, a so-called generalized Floquet Hamiltonian (originally defined by Jauslin and Lebowitz in \cite{JaL}), a self-adjoint operator $K$ on an enlarged Hilbert space ${\cal K} = L^{2}({\cal M},\mu) \otimes {\cal H}$, may be defined, where ${\cal M}$ is a compact metric space endowed with a probability measure $\mu$; it is a circle, a torus or the hull of an almost-periodic function in the three cases above. Defining
$$
U_{K}(t,t_{0}) \equiv \exp[-i(t-t_{0})K]
\eqno{(5.40)}
$$
the relations (5.32) become
$$
\exp[is\exp(\lambda_{j}(t-t_{0})K_{\alpha_{j}})] U^{\dag}(t,t_{0}) = U^{\dag}(t,t_{0}) \exp(isK_{\alpha_{j}})
\eqno{(5.41)}
$$
Above, we consider a multidimensional generalization of (5.35.1),
$$
H(t) = p^{2}/2 + [x^{T} A(t) x]/2
\eqno{(5.42)}
$$
where $x^{T} \equiv (x_{1}, \cdots x_{n})$, $p^{T} \equiv (p_{1}, \cdots p_{n})$, $T$ denoting the transpose of the column vectors ($x$ and $p$). $A$ is a real symmetric matrix depending almost-periodically on time (see remark 4 of \cite{SJWe}). In (5.42), $\lambda_{j}$ are $2n$ complex numbers such that
$$
\Re(\lambda_{1}) \le \cdots \Re(\lambda_{n}) < 0 < \Re(\lambda_{n+1}) \le \cdots \Re(\lambda_{2n})
\eqno{(5.43.1)}
$$
i.e., $\lambda_{2n}$ is the generalization of the upper quantum Lyapunov exponent defined as in (5.39) with $\mathbf{R}^{2}$ replaced by $\mathbf{R}^{2n}$; analogously (5.36) becomes
$$
K_{\underline{\alpha}} = \alpha^{T} x + \alpha_{p}^{T} p \mbox{ with } \underline{\alpha} \in \mathbf{R}^{2n}
\eqno{(5.43.2)}
$$
with $\underline{\alpha}$ denoting the column vector with components $(\alpha_{x},\alpha_{p})$, and (5.34) becomes
$$
W(\underline{\alpha}) = \exp[i(\alpha_{x}^{T} x + \alpha_{p}^{T} p)]
\eqno{(5.34.2)}
$$
with $\underline{\alpha} \in \mathbf{R}^{2n}$. Finally, (5.37) is replaced by
$$
[L_{\underline{\alpha}}, W(\underline{\alpha^{'}})] =
 -\sigma(\underline{\alpha},\underline{\alpha^{'}}) W(\underline{\alpha^{'}})\quad
\mbox{ for all } \underline{\alpha}, \underline{\alpha^{'}} \in \mathbf{R}^{2n}
\eqno{(5.37.1)}$$
with
$$
\sigma(\underline{\alpha},\underline{\alpha^{'}}) = \alpha_{x}^{T} \alpha_{p}^{'} - \alpha_{p}^{T} \alpha_{x}^{'}
\eqno{(5.37.2)}
$$
the usual symplectic form. We shall assume that the unstable eigendirection corresponding to, e.g.,
$(\alpha_{x}^{2},\alpha_{p}^{2})$ in the $\mathbf{R}^{2}$ - case be such that
$$
\alpha_{p 2} \ne 0
\eqno{(5.36.1)}
$$
or, in general, that
$$
\alpha_{p 2n} \ne 0
\eqno{(5.36.2)}
$$
This is always possible by a slight variation of $t_{0}$, see \cite{SJWe}. It is discussed in \cite{Th2} and \cite{JSGW} why (5.32), (5.42) are a natural quantum version of the classical concept of sensitive dependence on initial conditions.

We refer to (5.32) and (5.42) (with $\underline{\alpha} = \underline{\alpha}_{2n}$ corresponding  to $\lambda_{2n}$) as $(H,K)$ - case (a)- (resp. $(K,K_{\underline{\alpha}_{2n}})$ - case (b)-) Anosov systems.

Since Hamiltonians such as (5.31.1) and (5.35.1) are not semibounded (in the latter case we should more properly speak of the generalized Floquet operator), they are only an approximation to realistic systems, similarly to the Stark Hamiltonian for an atom in an external electric field, in which case the approximation may be excellent, under suitable physical conditions; as an example, (5.35.1) describes well the behavior of ions in so-called Paul traps \cite{FWW}.

We shall now introduce a class of models for decoherence, constructed from quantum Anosov systems as a starting point, in similar way as the Coleman-Hepp model in section 3.2.

\subsection{Quantum Anosov models of decoherence}

We adopt the general framework of section 3.5, although we shall be describing an interacting system, without necessarily identifying ${\cal I}$ to a measuring apparatus. In proposition 1 we remarked on the greater flexibility of definition 3. We now take for the system $S$ just one spin one-half, and ${\cal I}$ a particle (oscillator) described by a
$(H,K)$ or $(K,K_{\underline{\alpha}_{2}})$ - Anosov system (the general $n$ case is analogous). The Hamiltonian $H_{c}$ given by (3.20), with $H$ a multiple of $\sigma_{0}^{3}$, $R$ the first member of either a $(H,K)$ (case (a)) or
$(K,K_{\underline{\alpha}_{2}})$ Anosov system (case (b)), and
$$
V_{1} = \mu K \sigma_{0}^{3}
\eqno{(5.45.1)}
$$
in case (a), and
$$
V_{2} = \mu K_{\underline{\alpha}_{2}} \sigma_{0}^{3}
\eqno{(5.45.2)}
$$
in case (b). Since $H$ is a multiple of $\sigma_{0}^{3}$ it will not play any role in the dynamics and will be omitted (the choice $H = \sigma_{0}^{1}$ is nontrivial and may be treated by perturbation theory, but we shall not do it in this paper). The Hamiltonians are, thus:
$$
h_{1} = H + \mu K \sigma_{0}^{3}
\eqno{(5.46.1)}
$$
in case (a) and
$$
h_{2} = K + \mu K_{\underline{\alpha}_{2}} \sigma_{0}^{3}
\eqno{(5.46.2)}
$$
in case (b). The corresponding initial states of the compound system $S_{c}$ are described, respectively, by the wave vectors
$$
\Psi_{1,2}(0) = \Psi_{0} \otimes \Psi^{1,2}
\eqno{(5.47.1)}
$$
where
$$
\Psi_{0} = c_{+} |+)_{0} + c_{-} |-)_{0}
\eqno{(5.47.2)}
$$
and
$$
\Psi^{1} = \phi
\eqno{(5.47.3)}
$$
$$
\Psi^{2} = \mathbf{1} \otimes \phi
\eqno{(5.47.4)}
$$
with
$$
\phi \in C_{0}^{\infty}(\mathbf{R})
\eqno{(5.47.5)}
$$
on ${\cal H} \otimes {\cal K}^{1,2}$, with ${\cal H} = \mathbf{C}_{0}^{2}$ and ${\cal K}^{1} = L^{2}(\mathbf{R})$,
${\cal K}^{2} = L^{2}({\cal M},d\mu) \otimes L^{2}(\mathbf{R})$. By (5.46) and (5.47), the time evolutes of $\Psi_{1,2}(0)$ at time $t \ge 0$ (which we henceforth assume) are given by
$$
\Psi_{i}(t) = c_{+} |+)_{0} \otimes \exp[-it(H +\mu K)] \Psi^{i}
+ c_{-} |-)_{0} \otimes \exp[-it(H -\mu K)] \Psi^{i}
$$
$$ \mbox{ for } i=1,2
\eqno{(5.48)}
$$
We shall extend definition 3 in the following sense:

\textbf{Definition 5} The systems $S^{i}$, $i=1,2$, described by (5.46.1),(5.46.2) are said to display decoherence or reduction of the wave-packet if there exists a $0 < t_{0} < \infty $ (''critical'' or ''decoherence'' time) and real numbers
$d_{\pm}^{i}$ , $i=1,2$, such that
$$
E_{i}(A) \equiv (\Psi_{i}(t), A \Psi_{i}(t)) = d_{+}^{i} (+|_{0} A |+)_{0} + d_{-}^{i}(-|_{0} A |-)_{0}
\eqno{(5.49)}
$$
for all $A \in {\cal B}(\mathbf{C}_{0}^{2})$ and for all $t \ge t_{0}$, $i=1,2$.

Of course, a system $S$ of arbitrary finite dimension may be included, as in (3.24.1).
This means, we are now interested in decoherence of the state of the small system ${\cal B}(\mathbf{C}_{0}^{2})$ and interpret its behaviour as reaction on the interaction with an environment.

Definition 5 is implicitly adopted in most applications in theoretical physics (\cite{Zu1},\cite{Zu2},\cite{KJZu},\cite{BKZu}), i.e., it takes over (3.24.1) without assuming the remaining structure (which is, however, of crucial importance in the mixed case and in general statistical mechanics, see remark 3). This definition is of relevance when the ''pointer'' has few degrees of freedom: see, in particular, studies \cite{BKZu} which have shown that an environment with a single number of degrees of freedom but displaying chaotic dynamics can be far more effective at destroying quantum coherence than a heat bath with infinitely many degrees of freedom - e.g., ''quantum brownian motion'' \cite{UZ}.

The following two lemmas will play a major role. We state them for $(H,K)$- Anosov systems, but they are also applicable to $(K,K_{\underline{\alpha}_{2n}})$-Anosov systems using (5.42).

\textbf{Lemma 4} Let $(H,K)$ define a quantum Anosov system. Then:
$$
\{\exp(it H/n) \exp(it \mu K/n)\}^m =
$$
$$
=\exp[it(\exp(-\lambda t/n)+ \exp(-2\lambda t/n) \cdots + \exp(-m\lambda t/n))\mu K] \exp(imt H/n)
\eqno{(5.50.1)}$$
$$
\{\exp(-itH/n) \exp(it\mu K/n)\}^m =
$$
$$
= \exp(-itmH/n) \exp[i\mu Kt/n(\exp(-\lambda t/n)+ \exp(-2\lambda t/n) + \cdots +\exp(-m\lambda t/n)]
\eqno{(5.50.2)}
$$
for $m,n$ arbitrary positive integers. In addition,
$$e^{it(H+\mu K)}=e^{i\mu K(\frac{1-e^{-\lambda t}}{\lambda})} e^{iHt}$$
$$e^{-it(H-\mu K)}=e^{-itH} e^{-i\mu K(\frac{1-e^{-\lambda t}}{\lambda})}
\eqno{(5.50.3)}
$$

\textbf{Proof} We use induction on $m$. For $m=1$, (5.50.1) reduces to (5.32). Assume, now, (5.50.1) valid for $m=m_{0}$, and consider the l.h.s. of (5.50.1) for $m=m_{0}+1$. By the induction hypothesis it equals
\begin{eqnarray*}
\exp(itH/n)\exp(it\mu K/n) \exp[it(\exp(-\lambda t/n)+ \cdots +\exp(-m_{0}\lambda t/n))\mu K] \exp(im_{0}tH/n)\\
= \exp(itH/n) \exp[it\mu K/n(1+\exp(-\lambda t/n)+ \cdots +\exp(-m_{0}\lambda t/n)]\exp(im_{0}tH/n)
\end{eqnarray*}$$\eqno{(5.51)}$$

Applying (5.32) to (5.51), with $s=1+\exp(-\lambda t/n)+ \cdots +\exp(-m_{0}\lambda t/n)$, we obtain (5.50.1)
for $m=m_{0}+1$, concluding the proof. The exponents in (5.50.1) and (5.50.2) are Riemann sums for the integral
$\int_{0}^{t} \exp(\lambda u) du = \frac{1-\exp(-\lambda t)}{\lambda}$. Hence, by the strong continuity of the group
$\alpha \to \exp(i\alpha K)$, and taking $m=n \to \infty$ in (5.50.1), (5.50.2), we obtain (5.50.3) by the Trotter product formula (see, e.g., \cite{RSI}, pp.295-296). q.e.d. \, \qquad 

\textbf{Theorem 2}\, The system composed of a two-level system and a quantum
$(H,K)$ Anosov system in the initial state (5.47.1) decoheres exponentially fast for $t\rightarrow \infty, \lambda <0 $. If in addition $\Phi $ has compact support it decoheres in finite time in the sense of Definition 5, if $\Phi $ has compact support. The decoherence time depends on $\lambda $ and the localization of $\Phi $.

 If $\lambda >0$ it decoheres for $t\rightarrow \infty $ only if the support of $\phi $ satisfies $S<4\mu/\lambda.$ In this case the decoherence time is given by
$$
t_{01} = \frac{|\log 2|}{\lambda}
\eqno{(5.52.1)}
$$
under the assumption
$$
a_{1} \equiv \frac{S \lambda}{4\mu} < 1
\eqno{(5.52.2)}
$$
for the $(H,K)$ system, and
$$
t_{02} = t_{0}+ \frac{|\log 2|}{\Re \lambda_{2}}
\eqno{(5.52.3)}
$$
under the assumption
$$
a_{2} \equiv \frac{S \Re(\lambda_{2}) \exp(t_{0}\Re(\lambda_{2}))}{4\mu \alpha_{p 2}} < 1
\eqno{(5.52.4)}
$$
for the $(K,K_{\underline{\alpha_{2}}})$ system. Above $S$ denotes the support of the function $\Phi$ in (5.47.5), i.e, the smallest real number such that $\Phi(x) = 0 \mbox{ if } |x|>S$.

If however these conditions are not satisfied it does not decohere at all. Therefore the decoherence properties are time irreversible!

\textbf{Proof} In order to prove (5.49) for $(H,K)$ Anosov systems, it suffices, by (5.48), to prove that
$$
{\cal I} \equiv (\Phi, \exp[it(H+\mu K)] \exp[-it(H-\mu K)] \Phi) = 0
\eqno{(5.53)}
$$
for $t>t_{01}$ under assumption (5.52.2). Using (5.50.3) in (5.53) we obtain
$${\cal I} = (\phi, \exp[2i\mu K \frac{1-\exp(-\lambda t)}{\lambda}] \phi)
\eqno{(5.54)}
$$
We now use the explicit form (5.31.2) of the dilation operator, together with
$\exp[i(\alpha x + \beta p)] = \exp(i \alpha x) \exp(i \beta p) \exp(-i \alpha \beta/2)$ and the fact that
$\exp(i \beta p)$ acts as a translation by $\beta$ to obtain (5.53) for $t>t_{01}$  under assumption (5.52.2).
The proof of assertion (5.53) for $t>t_{02}$ under assumption (5.52.4) for $(K,K_{\underline{\alpha}_{2}})$ Anosov systems is the same, using the Trotter product formula with $t=t_{0}$ as initial time (see (5.40) and the remark after (5.36.2)). q.e.d.

\subsection{Remarks on the connection with the ideas of Peres and Zurek and collaborators. The dependence of decoherence on special parameter values in the purely quantum case: breakdown of universality}

\textbf{Remark 8}\, Interactions such as (5.45.1,2) generalize the interaction linear in the $x$ coordinate in quantum Brownian motion (\cite{UZ}, \cite{JZ}), by replacing it by the dilation operator $K_{\underline{\alpha}}$. The latter, as (5.37) shows, may be interpreted as a derivation in the direction of phase space determined by $(\alpha_{p},\alpha_{x})$, and is, thus, naturally suggested by classical mechanics, see also \cite{JSGW}.

\textbf{Remark 9}\, 

The phenomenon of decoherence, according to definition 5, may also be characterized in specific models, e.g., a particle in a double well or a caricature thereof as a two-level system (see, e.g., \cite{GSa}, \cite{WC} for examples), as instability of tunneling. We refer to \cite{Wightman1}, \cite{Wightman2} for lucid reviews, in particular the relation to the problem of the inversion line, shape and chiral superselection sectors associated to pyramidal molecules. This tunneling instability is \textbf{universal} in the semiclassical regime (outside of which we refer to as ''purely quantum case''), viz., it occurs for an arbitrarily small disturbance of the double well potential (and even localized far away from the minima of the potential): see \cite{GMS}. This phenomenon, dubbed by Simon ''flea on the elephant'' \cite{S4} plays a major role in the approach of Landsman and Reuvers \cite{LandReu} to the important issue of how to obtain just one outcome in a given measurement.

Theorem 2 provides a concrete realization of an idea of Peres \cite{Pe} and Zurek and collaborators (\cite{BKZu}, \cite{KJZu})- that two nearly identical systems, prepared in identical states, but obeying slightly different dynamics, i.e., subject to slightly different Hamiltonians $H$ and $H+\delta H$, will evolve into two different states, whose inner product decays exponentially in time, when $H$ is the quantization of a chaotic Hamiltonian - in a purely quantum context. There is, however, a proviso: conditions (5.52.2) and (5.52.4) do not hold for $\mu$ small, except if $\lambda$ or $\Re(\lambda_{2})$ is small - and, then, $t_{01}$ and $t_{02}$ will be correspondingly large. This breakdown of the universality present in the semiclassical domain, due to the dependence of the tunneling instability or decoherence on special parameter values - was also observed in the purely quantum proofs of tunneling instability (see, e.g., \cite{GSa}\cite{WC}), as well as in the Anderson transition (see, e.g., \cite{Wre} and references given there).

\section{Conclusion: irreversibility versus collapse}

In thermodynamics irreversibility is expressed as monotonic increase of entropy. This thermodynamical concept of entropy is believed to be related to the entropy of states defined both in classical and quantum theory, here especially by von Neumann. Under an automorphic time evolution this entropy is conserved. To explain an increase we can either include a suitable interaction with a surrounding or a kind of coarse graining becomes necessary, together with conditions on the initial state to explain the arrow of time. In quantum theory this coarse graining can be considered as a reduction of the algebra, that can also be interpreted as ignoring some unobservable quantum correlations. This corresponds to decoherence effects as they are also important in the theory of quantum measurements and formally appear in the same way. Entropy is related to our knowledge of the system. Thus models that appear in the theory of measurements can support also our explanations for the origin of the second law of thermodynamics.

However, decoherence in the theory of measurements should be interpreted with more care. In classical theory optimal knowledge of the system guarantees certainty in the outcome of a measurement. This is no longer true in quantum theory. Our ignorance on the outcome of the experiment is expressed in the increase of entropy in theorem 1. If however the measurement is performed and we have observed the position of the pointer, the probability of its position can only be evaluated if we repeat the measurement. In the individual observation it has a definite value. We may assume if e.g. we follow the interpretation of von Neumann, that the measured system is now in a state corresponding to the position of the pointer. Reduction of the wave function turns into the collapse of the wave function. We have gained knowledge though not necessarily in the individual case, but in the average the entropy is reduced (Lemma 3).

If we believe in the validity of the second law and also in the fact, that the occurrence of state collapse as it happens in a measuring process is not restricted to the observation by an observer who is not part of the physical system (compare \cite{Haag}, VII.1-VII.3 , see also \cite{Haag1}) we run into two competing effects: one is the increase of entropy corresponding to the fact that correlations become unobservable. The other is a decrease of entropy due to non automorphic events. Both effects yield irreversibility, but nevertheless they are not cooperative but contrary to one another. The fact that the second law holds up to any doubts tells us that these non automorphic events must be rare in comparison to the time scale that is relevant in thermodynamics.

\textbf{Acknowledgements} We are deeply indebted to Geoffrey L. Sewell for several enlightening remarks on previous drafts of this paper, in particular for his emphasizing the nonmonotonic character of the entropy growth in theorem 1 and referring us to the work of van Kampen, as well as for pointing out some physical inconsistencies in a previous version of the model in section 5.12. We are also grateful to the referee for several helpful remarks.

\bibliography{minhabiblio-t.bib}

\begin{thebibliography}{100}

\bibitem{Wightman1}
A.~S. Wightman.
\newblock In M.~Ioffredo F.~Guerra and C.~Marchioro, editors, {\em Proc. Int.
  Workshop on probability methods in mathematical physics}. World Scientific
  Singapore, 1992.

\bibitem{Wightman2}
A.~S. Wightman.
\newblock Superselection sectors: old and new.
\newblock {\em Nuovo Cim. B}, 110:751, 1995.

\bibitem{vN}
J.~von Neumann.
\newblock {\em Mathematical foundations of quantum mechanics}.
\newblock Princeton university press, 1955.

\bibitem{WhZu}
J.~A. Wheeler and W.~H. Zurek.
\newblock {\em Quantum theory and measurement}.
\newblock Princeton University Press, 1983.

\bibitem{Zu1}
W.~H. Zurek.
\newblock In J.~Perez-Mercader J.~Halliwell and W.~Zurek, editors, {\em
  Physical origins of time asymmetry}. Cambridge University Press, 1994.

\bibitem{Zu2}
W.~H. Zurek.
\newblock Decoherence, einselection and the quantum origins of the classical.
\newblock {\em Rev. Mod. Phys.}, 75:715, 2003.

\bibitem{Pe}
A.~Peres.
\newblock {\em Quantum theory: concepts and methods}.
\newblock Kluwer academic publishers, 1995.

\bibitem{Giu}
C.~Kiefer J. Kupsch I O.~Stamatescu D.~Giulini, E.~Joos and H.~D. Zeh.
\newblock {\em Decoherence and the appearance of a classical world in quantum
  theory}.
\newblock springer Berlin, 1996.

\bibitem{Schlossauer}
M.~Schlossauer.
\newblock {\em Decoherence and the quantum-to-classical transition}.
\newblock Springer Heidelberg Berlin, 2007.

\bibitem{LandReu}
N.~P. Landsman and R.~Reuvers.
\newblock A flea on {Schr\"{o}dinger's} cat.
\newblock {\em Found. Phys.}, 43:373--407, 2013.

\bibitem{ABN1}
R.~Balian A.~E.~Allahverdyan and T.~M. Nieuwenhuizen.
\newblock Understanding quantum measurement from the solution of dynamical
  models.
\newblock {\em Phys. Rep.}, 525:1--166, 2013.

\bibitem{ABN2}
R.~Balian A.~E.~Allahverdyan and T.~M. Nieuwenhuizen.
\newblock {C}urie-{W}eiss model of the quantum measurement process.
\newblock {\em Eur. Phys. lett.}, 61:452, 2003.

\bibitem{He}
K.~Hepp.
\newblock Quantum theory of measurement and macroscopic observables.
\newblock {\em Helv. Phys. Acta}, 45:237, 1972.

\bibitem{Sew1}
G.~L. Sewell.
\newblock On the mathematical structure of quantum measurement theory.
\newblock {\em Rep. Math. Phys.}, 56:271, 2005.

\bibitem{Sew2}
G.~L. Sewell.
\newblock Can the quantum measurement problem be resolved within the framework
  of {Schr\"{o}dinger} dynamics?
\newblock {\em Markov Proc. Rel. Fields}, 13:425, 2007.

\bibitem{WWEm}
B.~Whitten-Wolfe and G.~G. Emch.
\newblock A mechanical quantum measuring process.
\newblock {\em Helv. Phys. Acta}, 49:45, 1976.

\bibitem{Land}
N.~P. Landsman.
\newblock Algebraic theory of superselection sectors and the measurement
  problem in quantum mechanics.
\newblock {\em Int. Jour. Mod. Phys. A}, 6:5349, 1991.

\bibitem{Gua}
I.~Guarneri.
\newblock Irreversible behaviour and collapse of wave packets in a quantum
  system with point interactions.
\newblock {\em J. Phys. A Math. Theor.}, 44:485304, 2011.

\bibitem{Se3}
G.~L. Sewell.
\newblock {\em Quantum theory of collective phenomena}.
\newblock Clarendon Press, Oxford, 1986.

\bibitem{Hug}
N.~M. Hugenholtz.
\newblock In R.~F. Streater, editor, {\em Mathematics of Contemporary Physics}.
  Academic Press, 1972.

\bibitem{Maes2}
C.~Maes.
\newblock On the origin and the use of fluctuation relations for the entropy.
\newblock {\em Seminaire Poincar\'{e}}, 2:29--62, 2003.

\bibitem{deRJMN}
W.~deRoeeck, T.~Jacobs, C.~Maes, and K.~Netocny.
\newblock An extension of the {Kac} ring model.
\newblock {\em Jour. Phys. A}, 36:11547, 2003.

\bibitem{FrSch}
J.~Fr\"{o}hlich and B.~Schubnel.
\newblock Quantum probability theory and the foundations of quantum mechanics.
\newblock arXiv 1303.1484.

\bibitem{GLeb}
S.~Goldstein and J.~L. Lebowitz.
\newblock On the boltzmann entropy of non-equilibrium systems.
\newblock {\em Physica D}, 193:53--66, 2004.

\bibitem{Leb}
J.~L. Lebowitz.
\newblock Time-asymmetric macroscopic behavior: an overview.
\newblock In W.~L.~Reiter G.~Gallavotti and J.~Yngvason, editors, {\em
  Boltzmanns Legacy}, pages 63--87. Eur. Math. Soc., 2008.

\bibitem{Pen}
O.~Penrose.
\newblock Foundations of statistical mechanics.
\newblock {\em Rep. Progr. Phys.}, 42:1937, 1979.

\bibitem{Pen1}
O.~Penrose.
\newblock {\em Foundations of statistical mechanics}.
\newblock Oxford, Pergamon Press, 1970.

\bibitem{tHW}
D.~ter Haar and H.~Wergeland.
\newblock {\em Elements of Thermodynamics}.
\newblock Addison Wesley Publ. Co., 1966.

\bibitem{Gr}
R.~B. Griffiths.
\newblock In J.~Perez-Mercader J.~Haliwell and W.~Zurek, editors, {\em Physical
  origins of time asymmetry}. Cambridge University Press, 1994.

\bibitem{JPi1}
V.~Jaksic and C.~A. Pillet.
\newblock Mathematical theory of nonequilibrium statistical mechanics.
\newblock {\em Jour. Stat. Phys.}, 108:787, 2002.

\bibitem{JPi2}
V.~Jaksic and C.~A. Pillet.
\newblock {NESS} in quantum statistical mechanics: where are we after ten
  years?
\newblock IAMP News Bulletin, January 2011, available at the web.

\bibitem{LYPR}
E.~H. Lieb and J.~Yngvason.
\newblock The physics and mathematics of the second law of thermodynamics.
\newblock {\em Phys. Rep.}, 310:1--96, 1999.

\bibitem{Haag}
R.~Haag.
\newblock {\em Local quantum physics - Fields, particles, algebras}.
\newblock Springer Verlag, 1996.

\bibitem{Zu3}
W.~H. Zurek.
\newblock Basis of classical apparatus: into what mixture does the wave packet
  collapse?
\newblock {\em Phys. Rev. D}, 24:1516, 1981.

\bibitem{Lind1}
G.~Lindblad.
\newblock Entropy, information and quantum measurements.
\newblock {\em Comm. Math. Phys.}, 33:305, 1973.

\bibitem{NTh2}
H.~Narnhofer and W.~Thirring.
\newblock Algebraic {K} systems.
\newblock {\em Lett. Math. Phys.}, 20:231, 1990.

\bibitem{Em}
G.~G. Emch.
\newblock Generalized {K}-flows.
\newblock {\em Comm. Math. Phys.}, 49:191--215, 1976.

\bibitem{NTh1}
H.~Narnhofer and W.~Thirring.
\newblock In A.~Amann H.~Atmanspacher and U.~M\"{u}ller-Herrold, editors, {\em
  On quanta, mind and matter}. Kluwer academic publishers, 1999.

\bibitem{ENST}
G.~G. Emch, H.~Narnhofer, G.~L. Sewell, and W.~Thirring.
\newblock Anosov actions on noncommutative algebras.
\newblock {\em Jour. Math. Phys.}, 35:5582, 1994.

\bibitem{LRu}
E.~H. Lieb and M.~B. Ruskai.
\newblock Proof of the strong subadditivity of quantum mechanical entropy.
\newblock {\em J. Math. Phys.}, 14:1938, 1973.

\bibitem{ArSe}
H.~Araki and G.~L. Sewell.
\newblock {KMS} conditions and local thermodynamical stability of quantum
  lattice systems.
\newblock {\em Comm. Math. Phys.}, 52:103, 1977.

\bibitem{HK}
R.~Haag and D.~Kastler.
\newblock An algebraic approach to quantum field theory.
\newblock {\em Jour. Math. Phys.}, 5:548, 1964.

\bibitem{BRo1}
O.~Bratelli and D.~W. Robinson.
\newblock {\em Operator algebras and quantum statistical mechanics I}.
\newblock Springer, 1987.

\bibitem{Wil}
I.~F. Wilde.
\newblock Lecture notes on local quantum theory and operator algebras.
\newblock available from Ivan F. Wilde homepage.ntl.world.com.

\bibitem{MWB}
Domingos H.~U. Marchetti and Walter~F. Wreszinski.
\newblock {\em Asymptotic Time Decay in Quantum Physics}.
\newblock World Scientific, 2013.

\bibitem{LaRu}
O.~E. Lanford and D.~Ruelle.
\newblock Observables at infinity and states with short range correlations in
  statistical mechanics.
\newblock {\em Comm. Math. Phys.}, 13:194, 1969.

\bibitem{Wehrl}
A.~Wehrl M.~Guenin and W.~Thirring.
\newblock Introduction to algebraic thechniques.
\newblock Lectures given at theoretical seminar series CERN 68-69.

\bibitem{BRo2}
O.~Bratelli and D.~W. Robinson.
\newblock {\em Operator algebras and quantum statistical mechanics II}.
\newblock Springer, 2nd edition, 1997.

\bibitem{Stre}
R.~F. Streater.
\newblock The {Heisenberg} ferromagnet as a quantum field theory.
\newblock {\em Comm. Math. Phys.}, 6:233, 1967.

\bibitem{Rob}
D.~W. Robinson.
\newblock The statistical mechanics of quantum spin systems.
\newblock {\em Comm. Math. Phys.}, 6:151, 1967.

\bibitem{LiR}
E.~H. Lieb and D.~W. Robinson.
\newblock The finite group velocity of quantum spin systems.
\newblock {\em Comm. Math. Phys.}, 28:251, 1972.

\bibitem{NS}
B.~Nachtergaele and R.~Sims.
\newblock {L}ieb-{R}obinson bounds and the exponential clustering theorem.
\newblock {\em Comm. Math. Phys.}, 265:119--130, 2006.

\bibitem{GlKa}
J.~Glimm and R.~V. Kadison.
\newblock Unitary operators in {C}*-algebras.
\newblock {\em Pac. J. Math.}, 10:547, 1960.

\bibitem{Scha}
R.~Schatten.
\newblock {\em Norm ideals of completely continuous operators}.
\newblock Springer Berlin, 1960.

\bibitem{B}
J.~S. Bell.
\newblock On wave packet reduction in the {Coleman-Hepp} model.
\newblock {\em Helv. Phys. Acta}, 48:93, 1975.

\bibitem{NTh3}
H.~Narnhofer and W.~Thirring.
\newblock Galilei invariant quantum field theories with pair interaction - a
  review.
\newblock {\em Int. Jour. Mod. Phys. A}, 17:2937--2970, 1991.

\bibitem{NTh4}
H.~Narnhofer and W.~Thirring.
\newblock Quantum field theories with {Galilei-invariant} interactions.
\newblock {\em Phys. Rev. Lett.}, 64:1863, 1990.

\bibitem{BuLdV}
D.~Loss D.~Burkard and P.~di~Vicenzo.
\newblock Coupled quantum dots as quantum gates.
\newblock {\em Phys. Rev. B}, 59:2070, 1999.

\bibitem{LMi}
C.~M. Lockhart and B.~Misra.
\newblock Irreversibility and measurement in quantum mechanics.
\newblock {\em Physica A}, 136:47, 1986.

\bibitem{J1}
J.~M. Jauch.
\newblock {\em Foundations of quantum mechanics}.
\newblock Addison Wesley, 1968.

\bibitem{El}
R.~S. Ellis.
\newblock {\em Entropy, large deviations and statistical mechanics}.
\newblock Springer Berlin, 1985.

\bibitem{Lind2}
G.~Lindblad.
\newblock Expectations and entropy inequalities for finite quantum systems.
\newblock {\em Comm. Math. Phys.}, 39:111, 1974.

\bibitem{Lind3}
G.~Lindblad.
\newblock Completely positive maps and entropy inequalities.
\newblock {\em Comm. math. Phys.}, 40:147, 1975.

\bibitem{Gr1}
R.~B. Griffiths.
\newblock Consistent histories and the interpretation of quantum mechanics.
\newblock {\em J. Stat. Phys.}, 36:219--272, 1984.

\bibitem{GH}
M.~Gell-Mann and J.~B. Hartle.
\newblock In K.~K. Phua and Y.~Yamaguchi, editors, {\em Proceedings of the 25th
  international conference on high energy physics}. World Scientific Singapore,
  1991.

\bibitem{O}
R.~Omn\'{e}s.
\newblock {\em The interpretation of quantum mechanics}.
\newblock Princeton University Press, 1994.

\bibitem{Th4}
W.~Thirring.
\newblock The histories of chaotic quantum systems.
\newblock {\em Helv. Phys. Acta}, 69:706, 1996.

\bibitem{Araki}
H.~Araki.
\newblock Relative entropy of states of von {Neumann} algebras.
\newblock {\em Publ. RIMS Kyoto Univ.}, 11:809, 1976.

\bibitem{OP}
M.~Ohya and D.~Petz.
\newblock {\em Quantum entropy and its use}.
\newblock Springer Verlag, 1993.

\bibitem{Wehrl1}
A.~Wehrl.
\newblock General properties of entropy.
\newblock {\em Rev. Mod. Phys.}, 50:221, 1978.

\bibitem{K}
H.~Kosaki.
\newblock Interpolation theory and the {Wigner-Yanase-Dyson-Lieb} concavity.
\newblock {\em Comm. Math. Phys.}, 87:315, 1982.

\bibitem{Liebb}
E.~H. Lieb.
\newblock Convex trace functions and the {Wigner-Yanase-Dyson} conjecture.
\newblock {\em Adv. Math}, 11:267, 1973.

\bibitem{Zel}
S.~Zelditch.
\newblock In G.~Naber J.~P.~Fran\c{c}oise and T.~S. Tsun, editors, {\em Quantum
  ergodicity and mixing of eigenfunctions in Encyclopedia Mathematical
  Physics}. Academic Press NY, 2006.

\bibitem{NGvK}
N.~G. van Kampen.
\newblock In E.~G.~D. Cohen, editor, {\em Fundamental problems in statistical
  mechanics}. North Holland Amsterdam, 1962.

\bibitem{CJT}
C.~J. Thompson.
\newblock {\em Mathematical statistical mechanics}.
\newblock Macmillan, 1971.

\bibitem{ABL}
P.~G.~Bergmann Y.~Aharonov and J.~L. Lebowitz.
\newblock Time symmetry in the quantum process of measurement.
\newblock {\em Phys. Rev.B}, 134:1410, 1964.

\bibitem{Maes1}
C.~Maes.
\newblock Fluctuation relations and positivity of the entropy production in
  irreversible dynamical systems.
\newblock {\em Nonlinearity}, 17:1305, 2004.

\bibitem{Bu}
D.~Buchholz.
\newblock New light on infrared problems: sectors, statistics, spectrum and all
  that.
\newblock arXiv 1301.2516 v.2 (14-1-2013).

\bibitem{Na}
H.~Narnhofer.
\newblock In R.~Kotecky, editor, {\em Phase Transitions}. World Scientific
  Singapore, 1993.

\bibitem{NRo}
H.~Narnhofer and D.~W. Robinson.
\newblock Dynamical stability and thermodynamic phases.
\newblock {\em Comm. Math. Phys.}, 41:89, 1975.

\bibitem{CCF2}
R.~Carlone C.~Cacciapuoti and R.~Figari.
\newblock Decoherence induced by scattering: a three-dimensional model.
\newblock {\em J. Phys. A Math. Gen.}, 38:4933, 2005.

\bibitem{UZ}
W.~G. Unruh and W.~H. Zurek.
\newblock Reduction of the wave packet in quantum brownian motion.
\newblock {\em Phys. Rev. D}, 40:1071, 1989.

\bibitem{JZ}
E.~Joos and H.~R. Zeh.
\newblock The emergence of classical properties through interaction with the
  environment.
\newblock {\em Zeitschr. Physik B}, 59:223, 1985.

\bibitem{PuWo}
W.~Pusz and S.~Woronowicz.
\newblock Passive states and {KMS} states for general quantum systems.
\newblock {\em Comm. Math. Phys.}, 58:273, 1978.

\bibitem{MK}
W.~A. Majewski and M.~Kuna.
\newblock On quantum characteristic exponents.
\newblock {\em J. Math. Phys.}, 34:5007, 1993.

\bibitem{ViM}
R.~Vilela Mendes.
\newblock Lyapunov exponent in quantum mechanics. {A} phase space approach.
\newblock {\em Phys. Lett. A}, 171:253, 1992.

\bibitem{JSGW}
H.~R. Jauslin, O.~Sapin, S.~Gu{\' e}rin, and W.~F. Wreszinski.
\newblock Upper quantum {Lyapunov} exponent and parametric oscillators.
\newblock {\em Jour. Math. Phys.}, 45:4377, 2004.

\bibitem{SJWe}
O.~Sapin, H.~R. Jauslin, and S.~Weigert.
\newblock Upper quantum {Lyapunov} exponent and {Anosov} relations for quantum
  systems driven by a classical flow.
\newblock {\em Jour. Stat. Phys.}, 127:699, 2007.

\bibitem{Th2}
W.~Thirring.
\newblock What are the quantum mechanical {Lyapunov} exponents.
\newblock In P.~Urban, editor, {\em Proceedings of the 34 Internationale
  Universit{\" a}tswoche f{\" u}r Kern und Teilchenphysik Schladming}, pages
  223--237. Springer, 1996.

\bibitem{JaL}
H.~R. Jauslin and J.~L. Lebowitz.
\newblock Spectral and stability aspects of quantum chaos.
\newblock {\em Chaos}, 1:114, 1991.

\bibitem{FWW}
J.~Wu M.~Feng and K.~Wang.
\newblock A study of the characteristics of the wave packets of a {Paul}
  trapped ion.
\newblock {\em Commun. Theor. Phys.}, 29:497, 1998.

\bibitem{KJZu}
C.~Jarinski Z.~P.~Karkusevski and W.~H. Zurek.
\newblock Quantum chaotic environments, the butterfly effect, and decoherence.
\newblock {\em Phys. Rev. Lett.}, 89:170405, 2002.

\bibitem{BKZu}
R.~Blume-Kohout and W.~H. Zurek.
\newblock Decoherence from a chaotic environment: An upside-down ''oscillator''
  as a model.
\newblock {\em Phys. Rev. A}, 68:032104, 2003.

\bibitem{RSI}
M.~Reed and B.~Simon.
\newblock {\em Methods in modern mathematical physics - v.1, Functional
  Analysis}.
\newblock Academic Press, 1st edition, 1972.

\bibitem{GSa}
V.~Grecchi and A.~Sacchetti.
\newblock Critical metastability and destruction of the splitting in
  non-autonomous systems.
\newblock {\em Jour. Stat. Phys.}, 103:339, 2001.

\bibitem{WC}
W.~F. Wreszinski and S.~Casmeridis.
\newblock Models of two-level atoms in quasiperiodic external fields.
\newblock {\em Jour. Stat. Phys.}, 90:1061, 1998.

\bibitem{GMS}
G.~Jona Lasinio, F.~Martinelli, and E.~Scoppola.
\newblock New approach to the semiclassical limit of quantum mechanics. {I.
  Multiple} tunnelings in one dimension.
\newblock {\em Comm. Math. Phys.}, 80:223, 1981.

\bibitem{S4}
B.~Simon.
\newblock Semiclassical analysis of low-lying eigenvalues: the flea on the
  elephant.
\newblock {\em J. Func. Anal.}, 63:123, 1985.

\bibitem{Wre}
Walter~F. Wreszinski.
\newblock Progress in the mathematical theory of quantum disordered systems.
\newblock {\em J. Math. Phys.}, 53:123307, 2012.

\bibitem{Haag1}
R.~Haag.
\newblock On the sharpness of localization of individual events in space and
  time.
\newblock arXiv 1303.6431.

\end{thebibliography}
\bibliographystyle{unsrt}
\end{document}